\documentclass[useAMS,usenatbib,usegraphicx]{mn2e}

\usepackage{lscape} 
\usepackage{placeins}

\title[Massive galaxies are early types since $z\sim1$]{Early type galaxies have been the predominant morphological class for
massive galaxies since only $z\sim1$}
\author[F. Buitrago et al.]{Fernando Buitrago$^{1,2}$ \thanks{E-mail: fb@roe.ac.uk}, Ignacio Trujillo$^{3,4}$, Christopher J. Conselice$^{1}$, Boris H\"au\ss ler$^{1}$ \\
\\
\\
$^{1}$University of Nottingham, School of Physics \& Astronomy, Nottingham, NG7 2RD, U.K. \\
$^{2}$SUPA\thanks{Scottish Universities Physics Alliance}, Institute for Astronomy, University of Edinburgh, Royal Observatory, Edinburgh, EH9 3HJ, U.K. \\
$^{3}$Instituto de Astrof\'{i}sica de Canarias, V\'{i}a L\'{a}ctea $s\backslash n$, 38200 La Laguna, Tenerife, Spain \\
$^{4}$Departamento de Astrof\'{i}sica, Universidad de La Laguna, E-38205, La Laguna, Tenerife, Spain }

\newcommand {\gtrsim} {\ {\raise-.5ex\hbox{$\buildrel>\over\sim$}}\ }
\newcommand {\lesssim} {\ {\raise-.5ex\hbox{$\buildrel<\over\sim$}}\ } 
\begin{document}


\maketitle

\label{firstpage}

\begin{abstract}
Present-day massive galaxies are composed mostly of early-type objects. It is unknown whether
this was also the case at higher redshifts. In a hierarchical assembling scenario the
morphological content of the massive population is expected to change with time from disk-like
objects in the early Universe to spheroid-like galaxies at present. In this paper we have
probed this theoretical expectation by compiling a large sample of massive
($M_{stellar}\geq10^{11} h_{70}^{-2} M_{\odot}$) galaxies in the redshift interval 0$<$z$<$3.
Our sample of 1082 objects comprises 207 local galaxies selected from SDSS plus 875
objects observed with the HST belonging to the POWIR/DEEP2 and GNS surveys. 639 of our objects
have spectroscopic redshifts. Our morphological classification is performed as close as possible to the optical restframe according to the photometric bands available in our observations
both quantitatively (using the S\'ersic index as a morphological proxy) and qualitatively (by
visual inspection). Using both techniques we find an enormous change on the dominant
morphological class with cosmic time. The fraction of early-type galaxies among the massive
galaxy population has changed from $\sim$20-30\% at z$\sim$3 to $\sim$70\% at z=0.
Early type galaxies have been the predominant morphological class for
massive galaxies since only $z\sim1$.
\end{abstract}



\section{Introduction}
\label{sec:intro}

The present-day massive galaxy population is dominated by objects with early-type morphologies
(e.g. Baldry et al. 2004, Conselice et al. 2006). However, it is still unknown whether this  was also the case at
earlier cosmic epochs. Addressing this question is key in our understanding of the physical 
processes that drive galaxy evolution, as galaxy morphology is directly linked to the
evolutionary paths followed  by these objects. In fact, a profound morphological transformation
of the massive galaxy population is  expected within the currently most favoured galaxy
formation scenario, the hierarchical model. For massive galaxies this model predicts a rapid
formation phase at 2$<$z$<$6 dominated by a dissipational in-situ star formation  fed by  cold
flows (Oser et al. 2010; Dekel et al. 2009; Keres et al. 2005) and/or gas rich mergers
(Ricciardelli et al.  2010; Wuyts et al. 2010; Bournaud et al. 2011). At the end of this phase,
massive galaxies are expected to be more flattened and  disk-like than their lower redshift
massive counterparts (Naab et al. 2009).  After this monolithic-like formation phase, massive
galaxies are  predicted to suffer a period of intense bombardment by minor satellites (Khochfar
\& Silk 2006; Hopkins et al.  2009; Feldmann et al. 2010; Oser et al. 2010; Quilis \& Trujillo 2012) that may
eventually transform the original disk-like population into the predominant present-day
spheroid-like population.

Although the above scenario is very suggestive of a deep morphological transformation of the
massive galaxy population, there is no compelling observational evidence supporting this
idea. However, some recent works suggests that this could be the
case (e.g. Oesch et al. 2010, van der Wel et
al. 2011, Cameron et al. 2011, Weinzirl et al. 2011, Law et al. 2012). Probing this transformation is difficult from the observational point of view due
to the scarce number of massive galaxies at high-z. However, the advent of wide area and deep near
infrared surveys (e.g. Dickinson et al. 2003, Scoville et al. 2007, Conselice et al. 2011)
have opened up the possibility of exploring a large number of these galaxies up to high
redshifts. In this paper we address, to the best of our knowledge for the first time, the issue of the morphological
transformation of massive galaxies using a statistical  representative sample of nearly
$\sim$1000 galaxies with $M_{stellar}\geq10^{11} h_{70}^{-2} M_{\odot}$ obtained from  the SDSS
DR7 (z$\sim$0; Abazajian et al. 2009), POWIR/DEEP2 (0.2$<$z$<$2; Bundy et al. 2006, Conselice
et al. 2007) and GNS (1.7$<$z$<$3; Conselice et al. 2011) surveys. We  have already conducted a
morphological quantitative analysis of the above galaxies in previous papers (Trujillo et al.
2007; Buitrago et al. 2008) where we have provided clear evidence for a significant size evolution
for these objects since z$\sim$3. However, a visual classification of these galaxies and an analysis of their overall profile shape have been
missing. In this paper we take advantage of the combined power of the visual and quantitative
morphological analysis to explore how the morphologies of the massive galaxy
population has changed with redshift.

The structure of the paper is as follows: Section \ref{sec:data} is devoted  to the data
description and its analysis, Section \ref{sec:results} presents our main results, Section \ref{sec:change}
details the evolution with redshift of the various morphological classes and in
Section \ref{sec:discussion} we discuss the implications of our findings.  At the end of this work we include an Appendix
containing the simulations we have performed to test the accuracy of our structural parameter
determination in the GNS. Hereafter, we adopt a cosmology with $\Omega_m$=0.3, 
$\Omega_\Lambda$=0.7 and H$_0$=70 kms$^{-1}$ Mpc$^{-1}$. We use a Chabrier (2003) IMF and magnitudes are provided in AB system, unless otherwise stated.

\section{Data}
\label{sec:data}

To accomplish our objectives and to be statistically meaningful we need a large number of massive 
galaxies at all redshifts within our study. Ideally we would also like to study all of our galaxies at a similar
restframe wavelength range. This is the reason behind our choice of working with several different
surveys. The imaging for the local Universe galaxy reference sample was obtained using the SDSS
DR7 (Abazajian et al. 2009) although our sample was selected from the NYU Value-Added Galaxy
Catalog (DR6). This catalog includes single S\'ersic (1968) profile fits for $2.65\times10^{6}$
galaxies (Blanton et al. 2005), from which $1.1\times10^{6}$ galaxies have spectroscopic
information. Stellar masses come from Blanton \& Roweis (2007), which uses 
Bruzual \& Charlot (2003) models (hereafter BC03) and a Chabrier (2003)
IMF. We limited our work to all the massive ($M_{\star}\geq10^{11} h_{70}^{-2} M_{\odot}$)
galaxies with spectroscopic redshifts up to $z=0.03$. We have selected this redshift as an upper limit as it contains a
local sample with a number of objects ($\sim$200) similar to the number of galaxies we have in
our higher redshift bins. On doing this we assure they are all affected statistically in a
similar way. By selecting z=0.03 we also guarantee that these galaxies are retrieved from a
sample that is complete in stellar mass. One object in this local sample was rejected as we
discovered it was a stellar spike. Our final local sample contains 207 galaxies. We have used the
g-band imaging of SDSS to classify visually our local sample.

In the redshift range $0.2<z<2$ we utilised the Palomar Observatory Wide-field InfraRed (POWIR)/DEEP2 survey
(Bundy et al. 2006, Conselice et al. 2007, 2008). For this part of the analysis, we restricted ourselves to the HST ACS
I-band coverage in the Extended Groth Strip (EGS) (see Lotz et al. 2008). It covers 10.1$\times$70.5 arcmin$^2$, for a total area of 0.2 deg$^2$.  The EGS field
(63 Hubble Space Telescope tiles) was imaged with the Advanced Camera for Surveys (ACS) in the V(F606W, 2660s) and
I-band (F814W, 2100s). Each tile was observed in 4 exposures that were combined to produce a pixel scale of 0.03
arcsec with a Point Spread Function (PSF) of 0.12 arcsec Full Width Half Maximum (FWHM). 
Complementary photometry in the B, R and I bands was taken with the CFH12K camera at the CFHT 3.6-m telescope and in
the $K_{s}$ and J bands in a large programme (65 nights) with the WIRC camera at the Palomar 5-m telescope.
The depth reached is $I_{AB} = 27.52$ $(5 \sigma)$ for point sources, and about 2 magnitudes brighter for extended objects.
For the purpose of the present article, we have analyzed the I-band imaging, although we stress our sample is K-band selected. All the massive galaxies in our sample have K$_{Vega} < 20$, being virtually complete for these objects at $z < 2$ (Conselice et al. 2007, see Section 4.1 and Figure 3). Simulations in Trujillo et al. (2007) agree on the detection of these objects in our ACS imaging.

For the highest redshift bins we used the GOODS NICMOS Survey\footnote{http://www.nottingham.ac.uk/astronomy/gns/} (GNS; Conselice et al. 2011). The GNS is a large HST NICMOS-3 camera program of 60 pointings centered around massive galaxies at $z = 1.7-3$ at 3 orbits depth, for a total of
180 orbits in the F160W (H) band. Its total area is 43.7 arcmin$^2$. Each tile (52"x52", 0.203"/pix) was observed in six exposures that were combined to
produce images with a pixel scale of 0.1 arcsec, and a PSF of $\sim0.3$ arcsec FWHM. The limiting magnitude of our observations is H$_{AB}$=26.8 (5$\sigma$ detection in a 0.7" aperture). The massive galaxies were firstly identified using a series of selection criteria: Distant Red Galaxies from Papovich et al. (2006), IRAC Extremely Red Objects from Yan et al. (2004) and BzK galaxies from Daddi et al. (2007). There is no photometric method or combination of methods able to perfectly identify a mass selected sample. Our primary identification, however, finds nearly all massive galaxies that would have been identified in a pure photometric redshift sample, with the exception of apparently rare blue massive galaxies (Conselice et al. 2011). Mortlock et al. (2011) shows that the survey imaging is complete for $M_{\ast} \gtrsim 10^{9.25} M_{\odot}$ (corrected to a Chabrier IMF) at $z = 3$ according to both mass function drops in the number of objects and theoretical predictions from maximally old stellar populations.

\subsection{Redshifts and masses, and photometric uncertainties}
\label{subsec:uncertainties}

\subsubsection{POWIR/DEEP2}
\label{subsubsec:powir}

In the POWIR/DEEP2 survey, 795 massive galaxies were used, with 421 of them possessing spectroscopic redshifts (from Davis et al. 2003). There were 35 more massive galaxies in the parent sample, but they were
excluded as they are identified as AGN and hence they may skew our results. When spectroscopic information
was not available, photometric redshifts were calculated for the bright galaxies ($R_{AB} < 24.1$) using the ANNZ code (Collister \& Lahav 2004) 
and BPZ (Ben\'itez 2000) for the rest. Accuracy is $\delta z/(1+z) = 0.025$ for $z < 1.4$ massive galaxies, and $\delta z/(1+z) = 0.08$
for the others (Conselice et al. 2007). Masses were calculated with the method described in Bundy, Ellis \& Conselice (2005), Bundy et al. (2006), Conselice et al. (2007): fitting
a grid of model SEDs constructed from BC03, parametrizing star formation histories by $SFR \propto exp(-t/\tau)$ (the so-called tau-model; with $\tau$ randomly selected from a range between 0.01 and 10 Gyr, and the age of the onset star formation ranging from 0 to 10 Gyr),
with a range of metallicities (from 0.0001 to 0.05) and dust contents (parametrized by the effective V-band optical depth $\tau_{V}$, where we use values $\tau_{V}$ = 0.0, 0.5, 1.2). To analyze the impact of thermally pulsating-asymptotic giant branch (TP-AGB) stars emission, the same exercise was also performed with Charlot \& Bruzual (2007; private communication)
models, inferring slightly smaller masses ($\sim10\%$). Combining the total uncertainties with those of the photometric redshifts, errors in the masses could be as high as 
$\sim32\%$ for $z > 1.4$ galaxies (Trujillo et al. 2007).
The sample of massive galaxies selected from this survey
constitutes the largest sample of HST observed massive galaxies in this redshift range published to date.

\subsubsection{GNS}
\label{subsubsec:gns}

The GNS total sample consist of 80 massive galaxies. We used spectroscopic redshifts (11) when available (Barger et al. 2008, Popesso et al. 2009).
Photometric redshifts and masses take advantage of the extensive GOODS fields wavelength coverage (BVRIizJHK), and were obtained using standard multicolor stellar population fitting techniques (Buitrago et al. 2008, Bluck et al. 2009, Conselice et al. 2011), making similar assumptions as in the POWIR/DEEP2 survey about exponentially declining star formation histories, using BC03 and a Chabrier (2003) IMF. Stellar masses errors are typically 0.2-0.3 dex. Spectroscopic redshifts agree well ($\delta z/(1+z) \sim 0.03$) with photometric determinations (Buitrago et al. 2008). As our observations are probing the optical restframe (or bluer than this), possible effects by TP-AGB stars are minimized. This sample is the largest massive galaxies compendium (80 objects) at $1.7 < z < 3$ imaged with HST we are aware of.

\subsubsection{Spectroscopic sample}
\label{subsubsec:spec}

One of the aims of the present work is selecting a sample with the largest number of spectroscopic redshifts as possible. This is extremely challenging at $z > 1.5$, even for objects as luminous as massive (M$_{\ast} > 10^{11} M_{\odot}$) galaxies, see for example Cimatti et al. (2008). Their spectral features become very difficult to detect, specially when reaching the redshift desert, i.e., when the [OII] emission line surpasses the 1 $\mu m$ wavelength limit of optical spectrographs. Consequently, at $z < 1.3$ two thirds of our sample posses spectroscopic redshifts, while at higher redshift the number of detections drop to just 10 in POWIR/DEEP2 and 11 in GNS. This fact implies that the redshift bins above $z=1.5$ are affected largely by these redshift uncertainties. However, given the quality of our HST photometry and the width of our redshift bins we argue that this has not a crucial impact in our final results.

\subsection{Quantitative morphological classification based on the S\'ersic index \textit{n}}
\label{subsec:qualitative}

Once we selected the final sample of objects, the surface brightness distributions of all our galaxies were fit to a single S\'ersic (1968) model  convolved with the PSF of the images. The S\'ersic model has the following analytical form:

$$I(r)=I_{e}\exp \left\{ -b_{n} \left[ \left( \frac{r}{a_{e}} \right) ^{1/n}-1 \right] \right\}$$

\noindent where $I_{e}$ is the intensity at the effective radius, and a$_e$ is the effective radius along the semimajor axis
enclosing half of the flux from the model light profile. The
quantity $b_{n}$ is a function of the radial shape parameter $n$ (called the S\'ersic index), which
defines the global curvature of the luminosity profile, and is obtained 
by solving the expression $\Gamma (2n) = 2\gamma (2n,b_{n})$, where $\Gamma (a)$
and $\gamma (a,x)$ are, respectively, the gamma function and the incomplete gamma function.
The sizes we provide are circularized, $r_{e}=a_{e}\sqrt{1-\epsilon}$, with
$\epsilon$ the projected ellipticity of the galaxy.

We first estimated 
the apparent magnitudes and sizes of our galaxies using SExtractor
(Bertin \& Arnouts 1996), which were then fed as initial conditions to the GALFIT code (Peng et al. 2010).
GALFIT convolves
S\'ersic r$^{1/n}$ 2D models with the PSF of the images and determines
the best fit by comparing the convolved model with the observed galaxy surface
brightness distribution using a Levenberg-Marquardt algorithm to minimise the
$\chi^2$ of the fit.

Before we carried out our fitting we masked neighbouring galaxies, following the idea explained in H\"au\ss ler et al. (2007) and Barden et al. (2012). In the case of very close galaxies with overlapping 
isophotes, objects were fit simultaneously. Due to the point-to-point variation of the shape of
the camera PSF in our images we chose several (non-saturated) bright stars to
gauge the accuracy of our parameter estimations. 
The final values for the structural parameters of every galaxy are the mean of these independent runs (one per each star used as PSF). 
The catalogs are published as part of Trujillo et al. (2007; POWIR/DEEP2 sample) and Conselice et al. (2011; GNS sample).

In relation to the SDSS imaging, although the NYU
catalog already provides structural parameters obtained using one dimensional S\'{e}rsic fits to the galaxies, for the sake of
consistency with our methodology, we ran again GALFIT on the SDSS images of these galaxies to obtain structural parameters. 
It is known that the NYU catalog has a systematic underestimation of the S\'{e}rsic index, effective radius and total flux,
as it is reported in the simulations performed in Blanton et al. (2005) and in the appendix of Guo et al. (2009).
Our findings agree with this fact, as we find an offset of $26\pm2\%$ for the circularised effective radius values of our galaxies
and another $14\pm3\%$ for the S\'{e}rsic indices.

We performed simulations to assess the reliability of our quantitative results at high redshift. Our purpose was two-folded: correcting our structural parameters for a more accurate 
representation of their values and exploring in an unbiased way whether we are missing any region in the parameter space of the massive galaxy properties. In case of the POWIR/DEEP2 sample the simulations 
are fully explained in Trujillo et al. (2007). They created 1000 artificial galaxies with random structural properties within the observed ranges of the sample. 
In that study no significant trend in either the sizes or the concentration of the galaxies was found (see their Fig.
3), except for an underestimation of $\sim20\%$ in the S\'{e}rsic index of the very faint
$I_{AB} > 24$  spheroid-like galaxies. We carried out a similar analysis in Appendix A for the galaxies in the GNS sample. For the present paper, we went a step further in the complexity of our simulations, not only in the number of objects ($\times$16) but specially in finely sampling the small sizes regime ($0.15-0.3$ arcsec) which is key to understanding the compact massive galaxy population.

Here we offer a brief description of our simulations. We created 16000 mock galaxies employing the IDL routines developed for H\"au\ss ler et al. (2007), randomizing both the galaxy positions in our imaging and the galaxy structural parameters. Once the output of our simulations in the parameters space ($H_{\rm output},r_{\rm e,output},n_{\rm output}$) was recovered, we estimated the difference with the intrinsic (input) values. 
The most important outcome is that, on average, the S\'ersic indices of our observed GNS massive galaxies grow by $\sim$10\% when corrections based on the simulations are applied. However, the individual values per galaxy depend on its exact position in the 3D space defined by the apparent magnitude, S\'ersic index and effective radius.

In general, we found that for objects with disk-like surface brightness profiles
(i.e n$_{\rm input}$$<$2.5), both sizes and S\'ersic indices are recovered with basically no
bias down to our limiting explored H-band magnitude. However, by increasing the input
S\'ersic index we found biases in the determination of the sizes and $n$. For example, a galaxy with
n$_{\rm input}$$\sim$4 and H=22.5 mag (our median magnitude within the GNS catalogue), the
output effective radii are $\sim10\%$ smaller and output S\'ersic indices are $\sim15-20\%$
smaller than the input values. The results of these simulations, however, show that the decrease
in the S\'ersic index we observe from z$\sim$2.5 to z=0 for the spheroid-like population
(which is around a factor of $\sim$2) cannot be fully explained as a result of the bias on
recovering the S\'ersic index. 

\subsection{Qualitative visual morphological classification}
\label{subsubsec:qualitative}

\begin{figure*}
\vspace{0.5cm}
\hspace{2cm}
\rotatebox{0}{
\includegraphics[angle=90,width=1.2\linewidth]{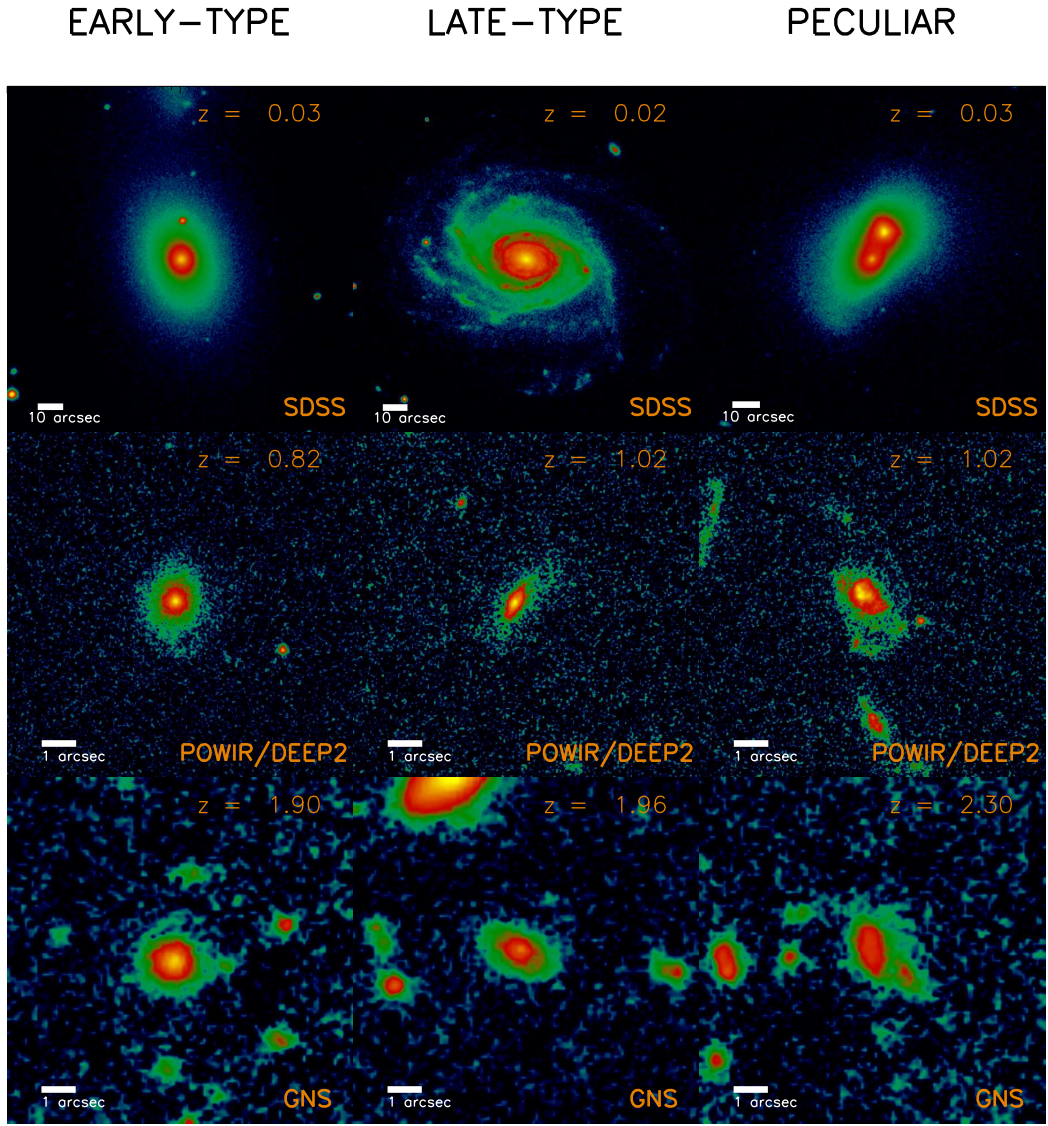}}
\vspace{-0.5cm}

\caption{Some examples illustrating our morphological criteria (columns) for different
galaxies in our sample. Each row shows galaxies from the different surveys. Please note the
different scales of each image due to the different redshift coverage of each survey (lower left corner); according
to the cosmology used in this work, 10 arcsec in SDSS are $\sim6$ kpc at z$\sim$0.03,
while 1 arcsec in the HST imaging at $z\geq1$ is $\sim8$ kpc. Despite the decrease in
angular resolution and the cosmological surface brightness dimming with redshift, the
exquisite HST depth and resolution ($\sim$10 times better than ground-based SDSS imaging)
allow us to explore the morphological nature of the high-z galaxies. Note that irregulars
and mergers are in the same morphological class (peculiars).}
\label{fig:montage}

\end{figure*}

In addition to the quantitative morphological analysis explained above, visual
morphological classifications were derived for all the galaxies in our sample. To assure a
high reliability in our results, two authors of this paper (FB and IT, with checks by CC) classified
visually all the galaxies in an independent way. We divided our sample according to the Hubble
classification scheme into spheroid-like objects (E+S0 or early-type), disk-like objects
(S or late-type) and peculiar galaxies (either irregular galaxies or ongoing mergers). In
Figure \ref{fig:montage} we show some examples of our classification scheme at different
redshifts. Very conspicuous bulge systems were identified as early-type objects. Both E
and S0 galaxies are hence included together in the same morphological class. We avoid
segregating between E and S0 since, at high-z, it is a  difficult task to distinguish
between these two types of galaxies, and we prefer
to  remove this potential source of error. Spiral or late-type morphologies are detected
by a central brightness condensation located at the centre of a thin disk containing more
or less visible spiral arms of enhanced luminosity. Lastly we joined irregular
(unsymmetrical) galaxies and mergers in the same class, again to avoid any
missclassification at high-z where the differences between these types are more difficult to interpret.

It is not straightforward to asses the robustness of our visual classifications as this is a pattern
recognition problem and as such cannot be addressed by standard algorithm procedures. Nevertheless, at z$\sim$0,
we can compare our results with other alternatives coming from independent studies. First, we compare our results with the SDSS
Bayesian automated morphological classification by Huertas-Company et al. (2011). There
are 190 out of our 207 galaxies in common where we can make a direct comparison.
They applied support vector machine techniques (Huertas-Company et al. 2008) to associate a
probability to each galaxy  being E, S0, Sab or Scd. For those galaxies where they have
assigned a probability larger than 90\% of pertaining to a given class, their neural
network agrees with our visual classification for 89\% of the early-types and 68\% of the
late-types. Moreover, all our SDSS local galaxies have been visually classified within the
Galaxy Zoo project (Lintott et al. 2011). We find that 112 out of the 121 galaxies that we
classified as early-type are classified as ellipticals by Galaxy Zoo (i.e. $\sim$93\%).
For spiral galaxies we get 48 out of 62 (i.e. $\sim$77\%). The discrepancies in our 
late-type galaxies identifications arise from the difficulty to interpret S0 galaxies without
the astronomer's ``trained-eye'' intervention and, as stated previously, they are included as early-type
objects throughout our study. Consequently, our local classification seems to be robust.

\subsection{Potential observational biases}
\label{subsec:pot_biases}

We acknowledge, however, that at higher redshifts visual morphological classification
is more controversial for several reasons. First, the cosmological surface brightness
dimming may affect the recognition of fainter galactic features and second, the angular
resolution is poorer at higher redshift. Nonetheless, the first effect is 
compensated by the increase of the intrinsic surface brightness (e.g. Marchesini et al. 2007; Prescott, Baldry \& James 2009, Cirasuolo et al. 2010) of the galaxies due to
having a higher star formation rate in the past, and the fact that their stellar
populations are younger. In relation to the angular resolution, at z=0.03, one arcsec is
equivalent to $\sim$0.6 kpc, whereas at 1$<$z$<$3 it is $\sim$8.0 kpc. Fortunately, the
higher resolution imaging used for exploring the morphologies of our high-z galaxies
(FHWM$\sim$0.1-0.3 arcsec) compared to the local ones (FWHM$\sim$1.0-1.5 arcsec)
alleviates this problem, although in general, a smoother surface brightness distribution
due to the worse resolution is expected. All these effects combined imply that at
higher redshifts there would be a larger number of featureless objects that visually would
be confused with early-type galaxies. We will show in the next section that this is the
opposite of what we find, ultimately giving stronger support to the results of this paper.

Another source of uncertainty comes from the objects catalogued as peculiar class. This classification could be seen as a miscellaneous box where we included galaxies which do not fulfill 
neither early-type nor late-type descriptions. Furthermore, their photometry may be compromised by means of their multiple components. For the sake of clarity, we include here the number of irregular massive galaxies in our redshift bins (see Table \ref{tab:structural_data}) which is 2/0/11/27/14/8, while for mergers are 5/8/31/31/16/13. The aim of this paper is not to constrain the nature of this peculiar class, and due to the multiple nature of some of their members their results are only tentative.

Finally, we must bear in mind that dust obscuration may alter galaxy observables depending on the geometry. Its main effect is usually an artificial size increment and S\'ersic index lowering due to the fact that central concentrations of dust flatten the flux profile from the observed galaxy (Trujillo et al. 2006). 
In reality, the real contribution depends on the amount of dust and its distribution within the galaxies at study (see Moll\"enhoff, Popescu \& Tuffs 2006, where the authors also present corrections for disks scalelenghts and central surface brightnesses). However, both Trujillo et al. (2007) (see their Figure 4 and Section 4.2) and Buitrago et al. (2008) (see their Section 4) reported only small changes in the structural parameters when comparing ACS and NICMOS observations. We shall come back to this issue in more detail in our next section.

\subsection{Morphological K-corrections}
\label{subsec:k_correction}

The K-correction effect is another potential source of error both within the quantitative and
visual morphological classification. We have selected our filters at each survey to
minimise this effect, and to observe the galaxies as close as possible to the optical
restframe. Nonetheless, our classification at 1.3$\lesssim$z$\lesssim$2 could be
compromised by using F814W as this filter is tracing the UV restframe of our targets. 

We explored how relevant this effect is by analysing the properties of 24 galaxies with z$<$2
in our POWIR/DEEP2 sample which also have H-band NICMOS imaging. In Trujillo et al. (2007)
we discussed the size difference between the optical and near-infrared for these
galaxies (their Fig. 4). We did not find a systematic bias, but a scatter of 32\%. With regards
to the S\'ersic index,  we found an offset of $30\pm9\%$ towards larger indices in the
H-band. The difference in the visual morphology between the I and
the H-bands shows that 19 galaxies ($79\pm8\%$) have the same morphology in the two filters, while
only 5 ($21\pm8\%$) are catalogued differently. Our errors for visual classifications are represented by the standard deviations for a binomial distribution. We found in ACS 13 ($54\pm10\%$) early-type galaxies (10 in NICMOS, $42\pm10\%$), 6 ($25\pm9\%$) late-type galaxies (10 in NICMOS, $42\pm10\%$) and 5 ($21\pm8\%$) peculiar galaxies (4 in NICMOS, $17\pm8\%$). 

In addition to this analysis of galaxies
in the EGS, we compared the difference between the I and H band morphologies for those
galaxies in the GNS with 1.7$<$z$<$2 (which is the redshift interval where our POWIR/DEEP2
and GNS sample overlap). We used the I-band ACS imaging of the GOODS fields (Giavalisco et
al. 2004). Postage stamp images for 20 common galaxies were retrieved from the RAINBOW
database (Barro et al. 2011). Our GALFIT analysis showed that the effective radius and 
the S\'ersic index are recovered without any significant offset, but with a large scatter
as in the aforementioned Trujillo et al. (2007). Regarding their visual morphologies, we found that 6 galaxies ($30\pm10\%$) were not possible to
classify reliably due to the few pixels that compose their image, most probably due to
dust obscuration (Buitrago et al. 2008, Bauer et al. 2011). For the remaining 14 galaxies, 11
($55\pm11\%$) have the same visual morphology, while for 3 galaxies ($15\pm8\%$) there is a difference.
For the detections in the ACS camera, 4 ($28\pm12\%$) are early-types (6 for their NICMOS counterparts, $43\pm11$),
6 ($43\pm11\%$) are late-types (5 in NICMOS, $36\pm13\%$) and 4 ($28\pm12\%$) are peculiars (3 in NICMOS, $21\pm11\%$). 

There are a number of studies which have explored the K-correction issue for massive galaxies. Toft et al. (2007) analyzed a sample of 41 galaxies with masses greater than 
$10^{10} M_{\odot}$ at $1.9 < z < 3.5$ having NICMOS and ACS imaging as well. They reported that half of their Distant Red Galaxies where marginally detected or even disappeared in restframe UV bands. 
In contrast, Cassata et al. (2011) reported a weak K-correction within their 563 similar-mass passive galaxies comparing ACS and WFC3 observations. 
Our sample is not as high redshift as Toft et al. (2007), it is more massive on average and in addition we deal a variety of massive galaxies instead of red and passive galaxies as in Cassata et al. (2011). 
Summarising, K-correction undoubtedly plays a role, which is hard to delimite in our sample. Nevertheless, our tests help us undertanding that visual morphologies appear to be robust 
against these changes.

\subsection{Axis ratios}
\label{subsec:axis_ratios}

\begin{figure*}
\begin{center} 
\hspace{1.25cm}
\rotatebox{0}{
\includegraphics[angle=90,width=0.45\linewidth]{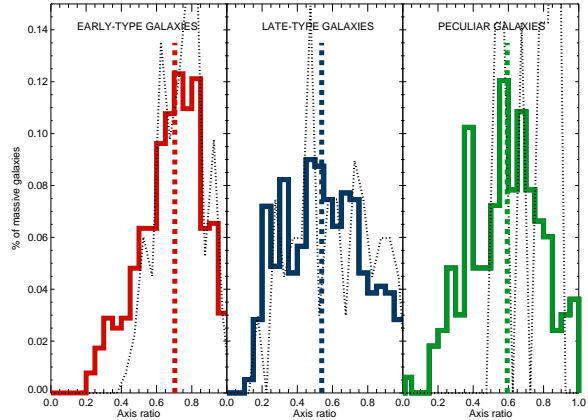}}
\end{center}
\vspace{-0.5cm}

\caption{Axis ratio distributions for our whole sample of massive galaxies according to their visual morphology. Vertical dashed lines are the median values of each histogram (0.70, 0.54 and 0.59, 
respectively). In order to have a local reference, we also overplotted our SDSS sample axis ratio distribution using the thin dotted line. 
The axis ratio distribution for the disks is rather symmetrical, but this is not the case for early-type objects, which peak at higher axis ratio values. Van der Wel et al. (2011) 
argued that massive galaxies with axis ratios $\lesssim$ 0.5 are most probably related with late-type objects, as this histograms confirm. Interacting/peculiar galaxies should be taken out 
of these considerations. The histograms are also in agreement with the estimations of axis ratio distributions for local early-type and late-type objects 
(Ryden, Forbes \& Terlevich et al. 2001; Ryden 2004). This fact highlights the reliability of our visual morphological classification.}
\label{fig:hist_ar}

\end{figure*}

Finally, we can conduct a further test to quantify the robustness of our visual
classification, namely to explore the axis ratio distribution of our objects. The axis ratio
distribution of local disk galaxies has a mean value of $\sim$0.5 (Ryden 2004). On the
other hand, the axis ratio distribution of the nearby E/S0 population is known to peak at
around 0.7-0.8 (Ryden, Forbes \& Terlevich et al. 2001). Figure \ref{fig:hist_ar} displays
the distributions of axis ratios in our sample according to our visual morphological determinations. 
We also overplot the median values from our samples (dashed line) and the axis ratio distribution using only our local sample (dotted line).
In Table \ref{tab:structural_data} we also show the mean axis ratio for our different galaxy populations. 
We find that the objects that are visually classified as
early-type galaxies have a typical axis ratio of $\sim$0.7 (independent of their
redshift). Also, for galaxies
visually classified as disks, the axis ratio is independent of the redshift with an average
b/a$\sim$0.55. Both values are in good agreement with the expectation from the local
Universe. This test reinforces the idea that our visual classifications are accurate.

\section{Results}
\label{sec:results}

The evolution of galaxy morphology with redshift can be addressed in two
different ways: quantitative (exploring how their structural parameters have changed with
time), and qualitative (probing how the visual appearance has evolved with redshift). In
the local Universe, the structural properties of massive galaxies (mainly its light
concentration) can be linked with their appearance. In particular, as a first
approximation one can identify disk or late-type galaxies with those galaxies having lower
values of the S\'ersic index (n$\sim$1; Freeman 1970) and early-type galaxies with those
having a profile resembling a de Vaucouleurs (1949) shape (n$\sim$4). This crude
segregation based on the S\'ersic index was shown to work reasonably well by Ravindranath
et al. (2004). Whether this equivalence also holds at higher redshift is not clear, and in
this paper we explore this issue.

\begin{figure*}
\rotatebox{0}{
\hspace{-2cm}
\includegraphics[angle=90,width=0.7\linewidth]{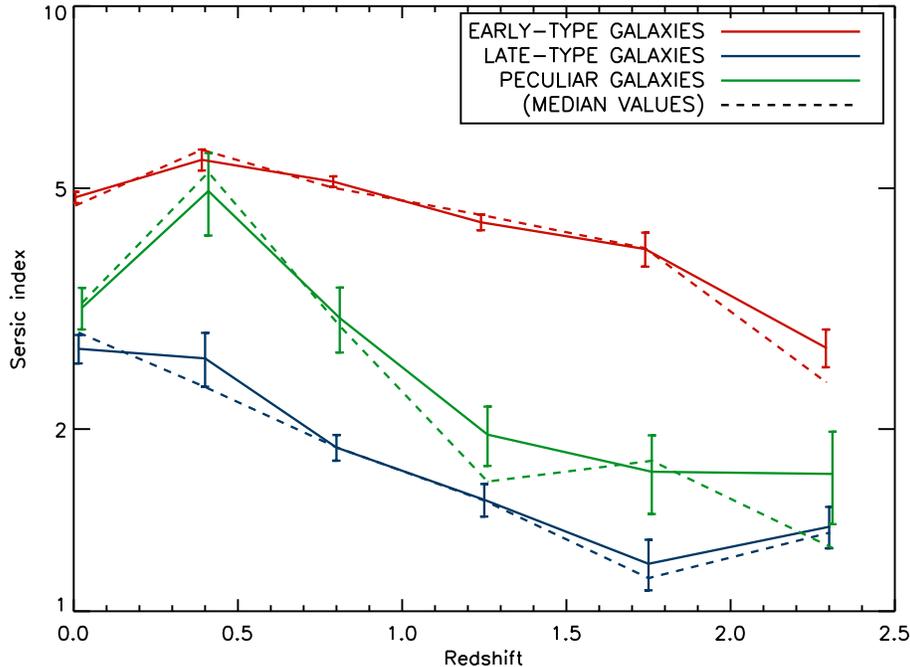}}

\caption{Evolution of the mean S\'ersic index values over redshift according the visual classifications of the galaxies within our sample (see also Table \ref{tab:structural_data}). Data points for early-types and peculiars are slightly offset for the sake of clarity. Dashed lines are similar but taking median values instead. Errors bars are the uncertainty of the mean ($\sigma/\sqrt{(N-1)}$, being $\sigma$ the standard deviation and N the total number of galaxies for each bin). They are slightly larger at 0.2$<$z$<$0.6 because of the comparatively poor statistics at this redshift interval. From that epoch to higher redshifts there is a clear separation between early-type massive galaxies and the rest of visual types, being all the average S\'ersic indices lower at increasing redshift.}
\label{fig:mean_vs_median}

\end{figure*}

Table \ref{tab:structural_data} lists the mean values of the structural parameters from the galaxies in our sample, splitting them among a number of redshift ranges and their visual morphologies. 
Concerning effective radius measurements, we retrieve again the reported size decrease for massive galaxies with redshift. As explained in Section \ref{subsec:axis_ratios}, axis ratios values per 
morphological class are independent of redshift, which agree with our morphological classifications. Finally, there is a tendency for the S\'ersic indices (observed using both mean and median values, 
see Fig. \ref{fig:mean_vs_median}) to become progressively smaller at increasingly higher redshifts. Moreover, there is a separation in the average Sersic index for late and early-type massive 
galaxies at all redshifts. Due to the number of photometric redshifts we used, it is interesting to know whether all these statistical tendencies are preserved if only taken into account
the results for galaxies with spectroscopic redshifts. This is in fact the case and it is presented in Table \ref{tab:structural_data_with_spec_z}.

\renewcommand{\arraystretch}{1.2}
\begin{center}
\begin{table*}
\caption{Observed mean structural parameters ($\pm$ the standard deviations) for visually classified massive (M$_{\star} > 10^{11}  h_{70}^{-2} M_{\sun}$) galaxies at 0$<$z$<$3}
\label{tab:structural_data}
\begin{tabular}{ccccccc}
\hline
\textbf{Early-type galaxies}  \\
\hline
Redshift Range & Number of galaxies & Survey & Effective radius & S\'{e}rsic index & Axis ratio & Mean stellar mass \\
  &   &   & (kpc) &   & (b/a) & ($10^{11} h_{70}^{-2} M_{\odot}$)  \\
\hline
0-0.03 & $133$ & SDSS & $ 7.15\pm1.56$ & $4.83\pm1.19$ & $0.74\pm0.13$ & $1.26\pm0.22$ \\
0.2-0.6 & $ 44$ & POWIR & $4.77\pm2.05$ & $5.57\pm1.46$ & $0.71\pm0.15$ & $1.51\pm0.45$ \\
0.6-1.0 & $184$ & POWIR & $3.52\pm1.79$ & $5.13\pm1.41$ & $0.67\pm0.19$ & $1.78\pm0.69$ \\
1.0-1.5 & $104$ & POWIR & $2.06\pm1.04$ & $4.39\pm1.32$ & $0.63\pm0.19$ & $1.70\pm0.51$ \\
1.5-2.0 & $ 30$ & POWIR & $1.31\pm0.73$ & $3.97\pm1.38$ & $0.65\pm0.17$ & $1.56\pm0.37$ \\
1.7-3.0 & $ 25$ & GNS & $1.30\pm0.55$ & $2.73\pm0.96$ & $0.68\pm0.11$ & $1.58\pm0.42$ \\
\hline
\hline
\textbf{Late-type galaxies} \\
\hline
Redshift Range & Number of galaxies & Survey & Effective radius & S\'{e}rsic index & Axis ratio & Mean stellar mass \\
  &   &   & (kpc) &   & (b/a) & ($10^{11} h_{70}^{-2} M_{\odot}$)  \\
\hline
0-0.03 & $ 67$ & SDSS & $ 8.44\pm3.28$ & $2.71\pm1.19$ & $0.60\pm0.22$ & $1.21\pm0.14$ \\
0.2-0.6 & $ 26$ & POWIR & $5.39\pm1.77$ & $2.62\pm1.28$ & $0.50\pm0.25$ & $1.40\pm0.30$ \\
0.6-1.0 & $124$ & POWIR & $4.91\pm2.04$ & $1.86\pm0.98$ & $0.54\pm0.21$ & $1.53\pm0.49$ \\
1.0-1.5 & $ 95$ & POWIR & $4.81\pm1.90$ & $1.53\pm0.87$ & $0.57\pm0.23$ & $1.58\pm0.41$ \\
1.5-2.0 & $ 42$ & POWIR & $3.88\pm1.51$ & $1.20\pm0.73$ & $0.50\pm0.20$ & $1.61\pm0.49$ \\
1.7-3.0 & $ 34$ & GNS & $2.55\pm1.18$ & $1.38\pm0.62$ & $0.54\pm0.18$ & $1.55\pm0.50$ \\
\hline
\hline
\textbf{Peculiar galaxies} \\
\hline
Redshift Range & Number of galaxies & Survey & Effective radius & S\'{e}rsic index & Axis ratio & Mean stellar mass \\
  &   &   & (kpc) &   & (b/a) & ($10^{11} h_{70}^{-2} M_{\odot}$)  \\
\hline
0-0.03 & $  7$ & SDSS & $ 8.39\pm2.22$ & $3.17\pm0.61$ & $0.72\pm0.13$ & $1.16\pm0.13$ \\
0.2-0.6 & $  8$ & POWIR & $4.93\pm2.26$ & $4.95\pm2.04$ & $0.56\pm0.23$ & $1.16\pm0.08$ \\
0.6-1.0 & $ 42$ & POWIR & $4.16\pm2.39$ & $3.05\pm2.40$ & $0.56\pm0.20$ & $1.65\pm0.49$ \\
1.0-1.5 & $ 58$ & POWIR & $3.83\pm1.64$ & $1.96\pm1.62$ & $0.61\pm0.18$ & $1.65\pm0.51$ \\
1.5-2.0 & $ 30$ & POWIR & $2.53\pm1.68$ & $1.70\pm1.36$ & $0.53\pm0.26$ & $1.81\pm0.68$ \\
1.7-3.0 & $ 21$ & GNS & $2.45\pm1.04$ & $1.69\pm1.31$ & $0.61\pm0.18$ & $1.44\pm0.34$ \\
\hline
\end{tabular}
\end{table*}
\end{center}
\renewcommand{\arraystretch}{1}

\renewcommand{\arraystretch}{1.2}
\begin{center}
\begin{table*}
\caption{Observed mean structural parameters ($\pm$ the standard deviations) for visually classified massive (M$_{\star} > 10^{11}  h_{70}^{-2} M_{\sun}$) galaxies with spectroscopic redshifts at 0.2$<$z$<$1.5}
\label{tab:structural_data_with_spec_z}
\begin{tabular}{ccccccc}
\hline
\textbf{Early-type galaxies}  \\
\hline
Redshift Range & Number of galaxies & Survey & Effective radius & S\'{e}rsic index & Axis ratio & Mean stellar mass \\
  &   &   & (kpc) &   & (b/a) & ($10^{11} h_{70}^{-2} M_{\odot}$)  \\
\hline
0.2-0.6 & $ 36$ & POWIR & $4.86\pm2.07$ & $5.73\pm1.51$ & $0.73\pm0.14$ & $1.51\pm0.46$ \\
0.6-1.0 & $124$ & POWIR & $3.54\pm1.88$ & $5.19\pm1.37$ & $0.68\pm0.19$ & $1.83\pm0.74$ \\
1.0-1.5 & $ 41$ & POWIR & $2.16\pm1.24$ & $4.61\pm1.34$ & $0.66\pm0.17$ & $1.91\pm0.61$ \\
\hline
\hline
\textbf{Late-type galaxies} \\
\hline
Redshift Range & Number of galaxies & Survey & Effective radius & S\'{e}rsic index & Axis ratio & Mean stellar mass \\
  &   &   & (kpc) &   & (b/a) & ($10^{11} h_{70}^{-2} M_{\odot}$)  \\
\hline
0.2-0.6 & $ 12$ & POWIR & $5.48\pm1.48$ & $2.90\pm1.22$ & $0.58\pm0.27$ & $1.49\pm0.30$ \\
0.6-1.0 & $ 68$ & POWIR & $5.09\pm2.11$ & $1.97\pm1.12$ & $0.60\pm0.21$ & $1.42\pm0.37$ \\
1.0-1.5 & $ 40$ & POWIR & $4.45\pm1.92$ & $1.46\pm0.81$ & $0.62\pm0.23$ & $1.56\pm0.38$ \\
\hline
\hline
\textbf{Peculiar galaxies} \\
\hline
Redshift Range & Number of galaxies & Survey & Effective radius & S\'{e}rsic index & Axis ratio & Mean stellar mass \\
  &   &   & (kpc) &   & (b/a) & ($10^{11} h_{70}^{-2} M_{\odot}$)  \\
\hline
0.2-0.6 & $  5$ & POWIR & $5.25\pm2.37$ & $5.12\pm2.40$ & $0.43\pm0.15$ & $1.16\pm0.09$ \\
0.6-1.0 & $ 22$ & POWIR & $3.79\pm2.11$ & $2.99\pm2.20$ & $0.60\pm0.21$ & $1.64\pm0.49$ \\
1.0-1.5 & $ 24$ & POWIR & $4.37\pm2.23$ & $1.38\pm1.13$ & $0.65\pm0.20$ & $1.88\pm0.70$ \\
\hline
\end{tabular}
\end{table*}
\end{center}
\renewcommand{\arraystretch}{1}

\begin{figure*}
\begin{center} 
\rotatebox{0}{
\includegraphics[width=0.70\linewidth,angle=90]{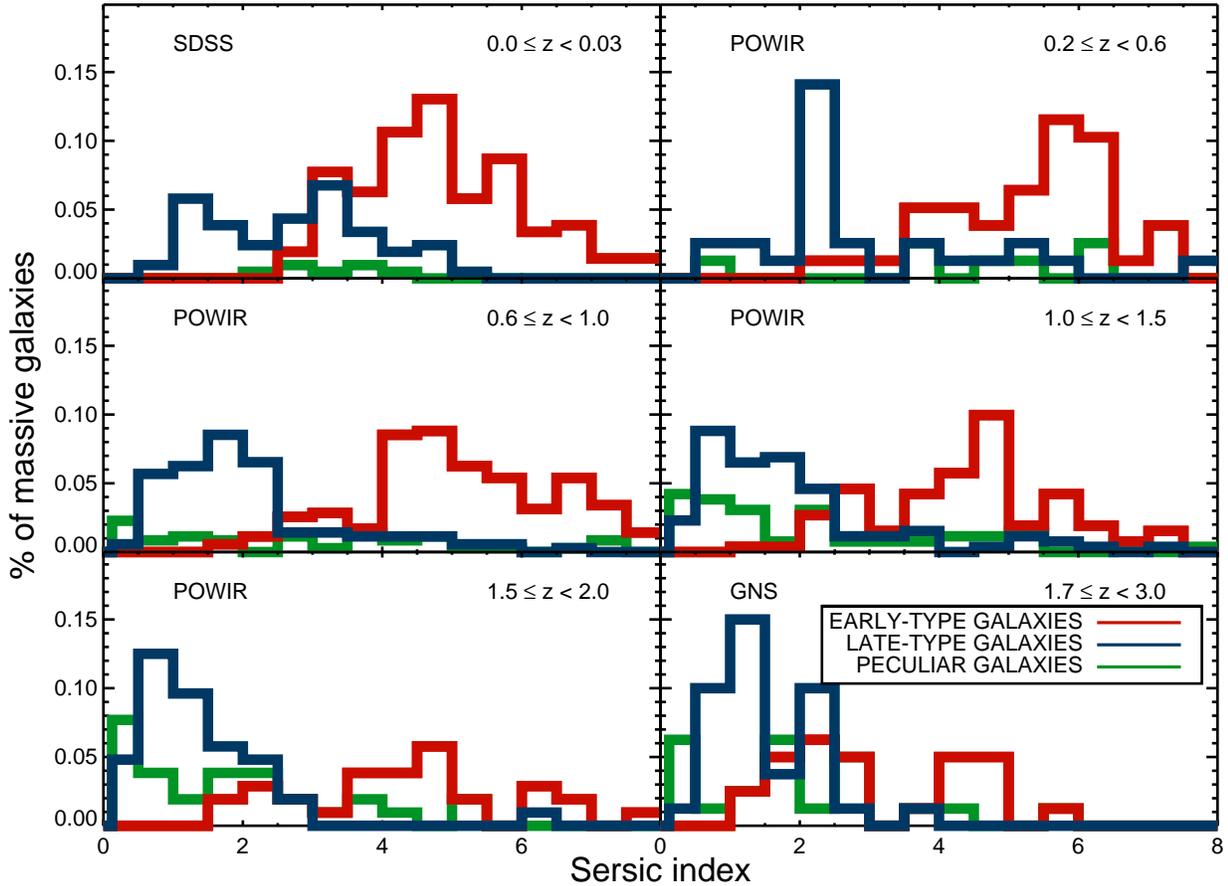}}
\end{center}

\caption{ S\'ersic index distribution of massive ($M_{\star}\geq10^{11} h_{70}^{-2}
M_{\odot}$) galaxies at different redshift intervals.  The S\'ersic indices of the
individual galaxies have been corrected following the simulations presented in Trujillo et
al. (2007; POWIR) and the Appendix of this paper (GNS). Color coding is related with visual morphology: blue for late-type galaxies, red
for early-type galaxies and green for peculiar (irregulars/mergers) galaxies. It is pertinent to note that the sharp peak for late-type objects at 0.2$<$z$<$0.6
is due to the small number statistics for this class of massive galaxies at this redshift interval (see Table \ref{tab:structural_data}). For our SDSS
sample, the S\'ersic index of disky objects are mainly located between 1$<$n$<$3 but for
some galaxies extend up to $n=5$. Conversely, the S\'ersic index of spheroid galaxies
starts at n$\sim$3 and then peaks at n$\sim$5.
This situation differs from our high-z results, where the very high S\'ersic indices have disappeared, and on average the values become smaller for all kinds of massive galaxies.} 
\label{fig:histograms}

\end{figure*}

\begin{figure*}
\rotatebox{0}{
\includegraphics[angle=90,width=0.95\linewidth]{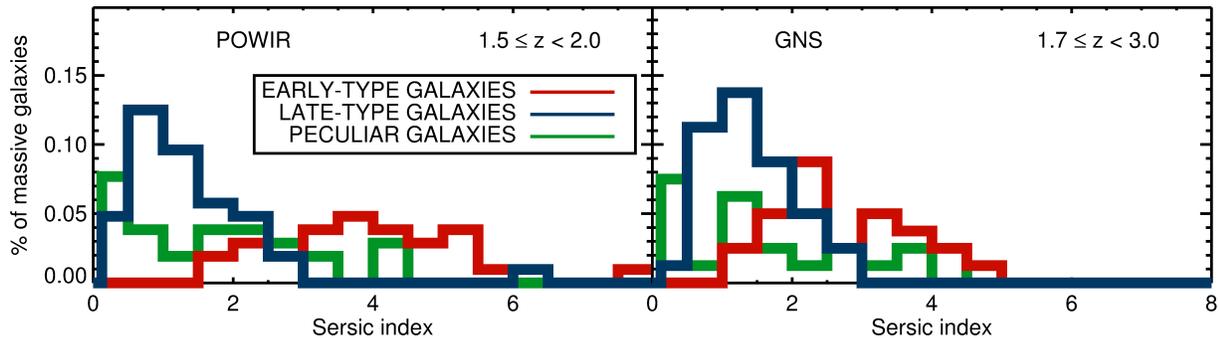}}

\caption{These are the highest redshift histograms of the Figure \ref{fig:histograms}, showing the observed S\'ersic indices values, without any a posteriori correction based on Trujillo et al. (2007) or our current GNS simulations (Appendix A). The more noticeable change is seen for the GNS data, where it is very conspicuous the non-existence of any large ($n > 4.62$) S\'ersic index. The difference between these histograms and the ones presented in Figure \ref{fig:histograms} is small.}
\label{fig:histograms_without_corrections}

\end{figure*}

We show in Fig. \ref{fig:histograms} the S\'ersic index distribution for our different visually classified
morphological types as a function of redshift. It is noteworthy that we had applied to this figure and to the next ones (unless otherwise stated) the corrections for the GNS (see Appendix A), 
and also for the POWIR/DEEP2 data using Trujillo et al. (2007) simulations. They only affect the two highest redshift bins. For the sake of
clarity, we also display in Figure \ref{fig:histograms_without_corrections} these histograms without any corrections. We do not appreciate that our corrections based on simulations affect the main results of our paper.

Coming back to Figure \ref{fig:histograms}, massive galaxies identified
visually as late-types show low S\'ersic index values at all redshifts. This reinforces the idea
that the stellar mass density distributions of rotationally supported systems are close to
an exponential profile. However, the distribution of the S\'ersic index for these
late-type galaxies shows a tail towards larger values. This is normally interpreted as the
result of the bulge component. In fact, the excess of light caused by the bulge at the
centre of the disk will increase the value of this concentration parameter when the
galaxies are fitted just using a single S\'ersic model. Interestingly, we observe that at
higher redshift the prominence of this tail of higher S\'ersic indices decreases for the
late-type galaxies. 

One could be tempted to interpret this result as a consequence of the
disappearance of prominent bulges at higher redshifts. However, a full exploration of this bulge development issue is beyond the scope of this paper. In the same Figure \ref{fig:histograms}, we show the distribution
of the S\'ersic index for massive galaxies visually classified as early-types. We see that
at low redshift, the distribution of S\'ersic indices for these galaxies predominantly show
large values of concentration as expected. Up to z$\sim$1.5 there is a peak around
n$\sim$4-6 (see also Table \ref{tab:structural_data}). A general trend is also observed: there is a progressive
shift towards lower and lower S\'ersic index values as redshift increases. 

\begin{figure*}
\vspace{1cm}
\rotatebox{0}{
\includegraphics[angle=0,width=1.\linewidth]{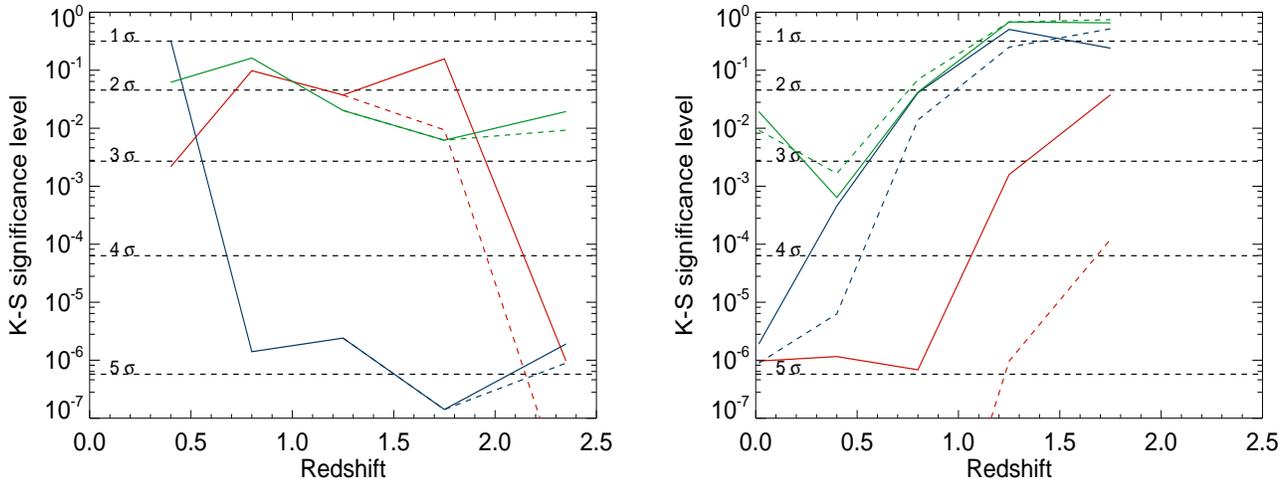}}

\caption{Kolmogorov-Smirnov significance levels comparing the
SDSS local sample S\'ersic index distribution with all the
other redshift bins (left panel) and the GNS highest redshift bin
S\'ersic index distribution against the rest of the distributions
(right panel). The color coding is the same as in the previous
figures: red is for early-type galaxies, blue is for late-type
galaxies
and green is for peculiar galaxies. Dashed lines represent the same
results but taking into account the corrections for S\'ersic
indices inferred from our simulations. The reason why dashed lines
encompass the whole plot in the right panel is that
the last distribution (GNS) is available with and without corrections,
but this does not occur for the local SDSS sample. The general
trends show that, within our uncertainties, the distributions are
diverging from the local relation (left panel) or progressively
converging up to the highest redshift bin (right
panel). The distributions with the S\'ersic indices corrected
according to our simulations show a lower departure with respect to
the fiducial
distribution, but still a significant one.}
\label{fig:k_s_tests}

\end{figure*}

We conducted a series of Kolmogorov-Smirnov (KS) tests in order to check the significance of the changes in the S\'ersic index with redshift. 
In Figure \ref{fig:k_s_tests} we show the evolution of the KS significance levels for the different morphological types in our sample of massive galaxies. 
The KS test is a non-parametric method of comparing probability distributions. We used as a base comparison the lowest redshift bin S\'ersic index distribution (left panel), 
and at the highest redshift one (right panel). Despite the uncertainties, mainly due to the number statistics, it seems that the distributions of S\'ersic indices gradually change from low to 
high redshift, and vice versa. When we conduct the analysis using the high-z bin as the reference instead of the local one, the evolution of the S\'ersic index distribution is statistically more 
significant on average. The fact that several redshift bins repeat significance level values in the left panel shows us the S\'ersic index distributions of early types are similar to the 
one at low-z, while this is not true for late-types. Possible explanations are the homology of early-type galaxies, and the double-peak which only exists in our local late-types S\'ersic indices, 
that is linked with the growth in the light profile wings. 

Consequently, the checks we carried out attempting to characterise the change of Sersic indices with redshift are positive 
(namely the Table \ref{tab:structural_data}, the Figure \ref{fig:mean_vs_median} and these KS tests), even though the KS comparison with the local S\'ersic index distribution is not as smooth as 
the others. The reason for this change or shift we observe could be either a real effect, produced by a decrease in the tail of the
surface brightness distribution of the massive galaxies at higher redshift, or an artificial one, produced by a bias at recovering large S\'ersic index values.

Again, our simulations were conducted for exploring potential observational artifacts. By means of comparing Figures \ref{fig:histograms} and \ref{fig:histograms_without_corrections}, 
one can check that the trend we observe towards lower S\'ersic indices at
higher redshifts is maintained with and without corrections. In fact, these corrections are minor. The only noticeable variation would be in the GNS early-types mean S\'ersic
index value in Table \ref{tab:structural_data}, which would grow up from 2.73$\pm$0.96 to 3.15$\pm$1.26. However, cosmic variance may partly be responsible for changing these numbers, 
as the GNS area (even though it was optimized for increasing the observed number of massive galaxies) is significantly smaller than the EGS POWIR/DEEP2 observations. 
Lastly, in relation to the distribution of the S\'ersic indices in Fig. \ref{fig:histograms} for the galaxies we classified as interacting or irregulars, we see a larger spread.
We discuss and interpret the S\'ersic index evolution for all our sample in the next section.

\section{The overall change in the massive galaxy population with redshift}
\label{sec:change}

\begin{figure*}
\vspace{1cm}
\rotatebox{0}{
\includegraphics[angle=0,width=1.\linewidth]{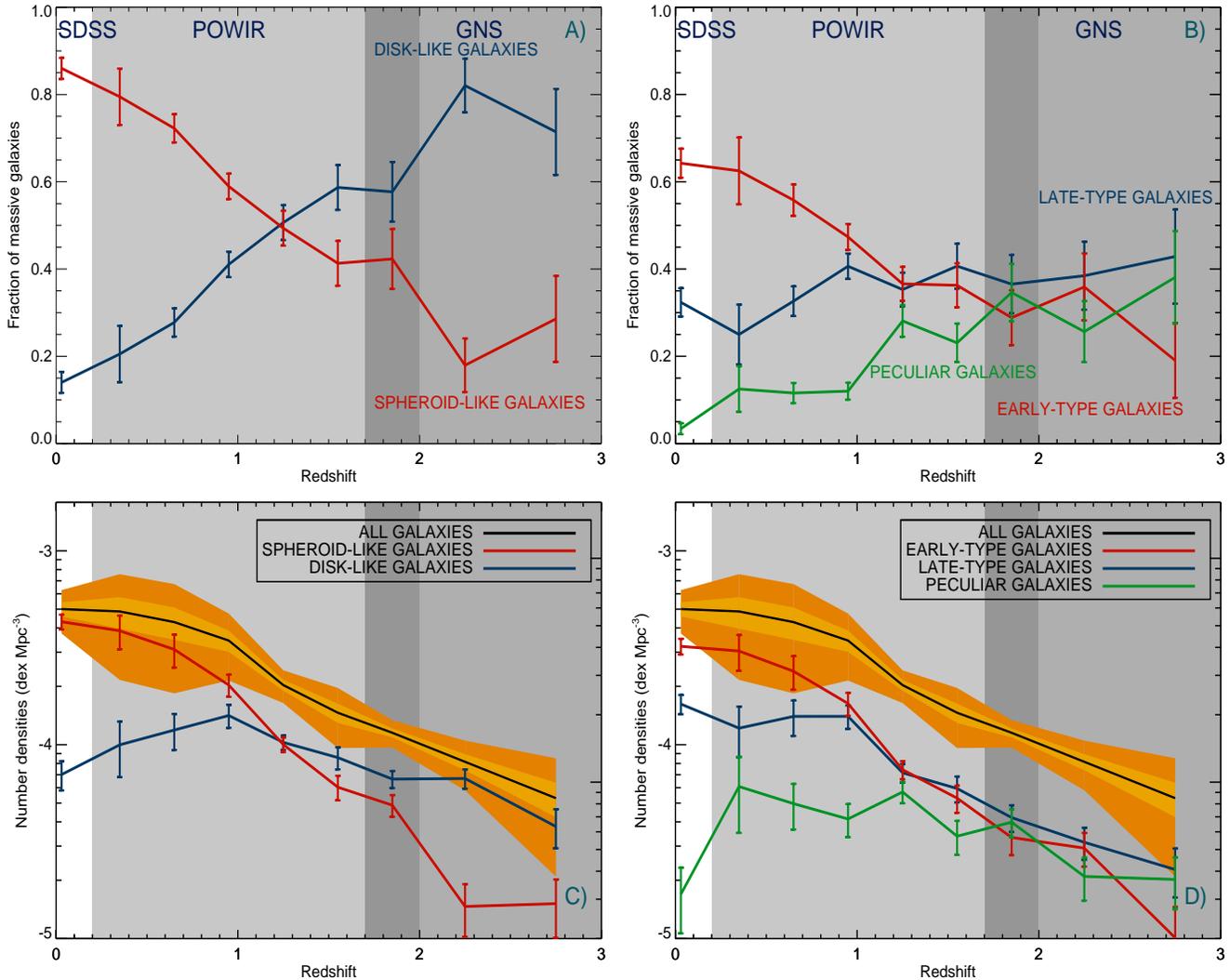}}

\caption{Panel A): Fraction of massive ($M_{*}\geq10^{11} h_{70}^{-2} M_{\odot}$)
galaxies showing disk-like surface brightness profiles ($n < 2.5$) and spheroid-like ones
($n > 2.5$) as a function of redshift. Different color backgrounds indicate the redshift
range expanded for each survey: SDSS, POWIR/DEEP2 and GNS. Error bars are estimated
following a binomial distribution. S\'ersic indices are corrected based on our simulations 
(Trujillo et al. 2007 and Appendix A in the present paper). B): Same as Panel A) but segregating the massive
galaxies according to their  visual morphological classification. Blue color represents
late type (S) objects and red early type (E+S0) galaxies, while
peculiar (ongoing mergers and irregulars) galaxies are tagged in green.  Panel C): Comoving number
density evolution of massive galaxies splitted depending on the S\'ersic index value. The
solid black line corresponds to the total number densities (the sum of the different components), with yellow and orange contours indicating
1$\sigma$ and 3$\sigma$ uncertainties in their calculation.} Panel D): Same as
panel C) but segregating the massive galaxies according to their visual morphological
type.
\label{fig:percentages}

\end{figure*}

Many studies (e.g. Shen et al. 2003; Barden et al. 2005; McIntosh et al. 2005; Trujillo et
al. 2006) have used n=2.5 as a quantitative way to segregate between early and late-type
galaxies. We explore, using this criteria, how the  percentages of the different types
of massive galaxies evolve with redshift. This is shown in Fig. \ref{fig:percentages}a. This figure clearly
indicates that the fraction of massive galaxies with lower S\'ersic index values has
dramatically increased at higher redshift. If the association between the S\'ersic index
and the global morphological type that holds at low redshift also applies at
high-z this would imply that massive galaxies at the high-z universe were mostly late-type
(disk) galaxies. However, there is no guarantee that such an association holds at all
redshifts. For this reason, we explore the evolution of the fraction of different
galaxy types with redshift using the visual morphologies (see Fig. \ref{fig:percentages}b). We find that the
population of visually classified massive disk galaxies remains almost constant with redshift
with perhaps a slight (if any) increase. The most dramatic changes are associated with the
early-type and irregular/mergers classes. The fraction of visually classified E/S0
galaxies increased by a factor of three since z$\sim$3 to now, whereas a reverse
situation is seen for the irregular/merging galaxies. This latter fact agrees with merging
becoming more important in massive galaxy evolution at increasing redshift (Conselice et al. 2009,
Bluck et al. 2009, 2012, L\'opez-Sanjuan et al. 2012). At z$\sim$2.5, late-type and peculiar objects account for the majority
of massive galaxies. One of the most important
outcomes of Fig. \ref{fig:percentages} is that the E/S0 type has been the dominant morphological fraction of
massive galaxies only since z$\sim$1. 

The number density of massive galaxies has significantly changed since z$\sim$3 (e.g.
Rudnick et al. 2003; P\'erez-Gonz\'alez et al. 2008a, Mortlock et al. 2011, Conselice et al. 2011) with a
continuous increase in the number of these objects in the last $\sim$11 Gyr. In order to
probe the emergence of the different galaxy types explored in this paper we have estimated
the comoving number density evolution of each class. To do this, we have used the Schechter fits to the stellar mass functions spanning from low to high redshift provided in a series of works: P\'erez-Gonz\'alez et al. (2008a) [$0 < z < 3$], Cole et al. (2001) [z = 0], Ilbert et al. (2010) [$0.2 < z < 2$], Kajisawa et al. (2009) [using the GALAXEV library (from BC03) alone and in combination with the EAZY code (Brammer et al. 2008); $0.5 < z < 3$] and Marchesini et al. (2009) [$1.3 < z < 3.0$]. All the mass functions utilized have been derived using BC03 models and normalized to our Chabrier IMF when necessary to be consistent with our data. We
have integrated these functions for all massive objects with $M_{stellar}\geq10^{11}
h_{70}^{-2} M_{\odot}$. The total number density for massive galaxies appears in Fig. \ref{fig:percentages}c \& \ref{fig:percentages}d as the black line. The yellow and orange backgrounds correspond to its 1$\sigma$ and 3$\sigma$ uncertainties. We have later multiplied those numbers by the fractions we have
estimated for the different classes of galaxies explored in this work. We show the
comoving number density evolution in Fig. \ref{fig:percentages}c \& \ref{fig:percentages}d, and the data is tabulated in Table \ref{tab:number_densities}. The number density of both disk-like and
spheroid-like massive galaxies, according to their S\'ersic index, have changed with time.
This evolution is particularly significant for spheroid-like objects which are now a
factor of $\sim$10 more numerous per unit volume than at z$\sim$2. About visual morphologies, the number of massive
late type galaxies has also increased as cosmic time progresses, but at a lower rate than early type
galaxies. Finally, the comoving number density of massive irregular/merging galaxies has grown
only very midly, if at all, in the last $\sim$11 Gyr.

\renewcommand{\arraystretch}{1.2}
\begin{center}
\begin{table*}
\caption{Comoving number densities for massive ($M_{\star}\geq10^{11} h_{70}^{-2} M_{\odot}$) galaxies (in $10^{-5} h_{70}^{3}$ dex Mpc$^{-3}$ units) }
\label{tab:number_densities}
\begin{tabular}{c|c|cc|ccc}
\hline
Redshift & $\Phi_{\rm total}$   & $\Phi_{\rm n<2}$ & $\Phi_{\rm n>2}$ & $\Phi_{\rm late}$ & $\Phi_{\rm early}$ & $\Phi_{\rm peculiar}$ \\
\hline
  0.03 & 50.07 $\pm$ 4.21 & 7.01   $\pm$ 1.21  & 43.06 $\pm$ 3.77 &   16.21   $\pm$  1.84 &   32.17 $\pm$ 2.99 &  1.69 $\pm$  0.63     \\   
  0.35 & 48.67 $\pm$ 9.03 & 9.98   $\pm$ 3.18  & 38.69 $\pm$ 7.63 &   12.17   $\pm$  3.55 &   30.42 $\pm$ 6.38 &  6.08 $\pm$  2.57     \\  
  0.65 & 42.89 $\pm$ 8.16 & 11.90  $\pm$ 2.51  & 30.99 $\pm$ 5.99 &   14.00   $\pm$  2.90 &   23.93 $\pm$ 4.71 &  4.97 $\pm$  1.31     \\
  0.95 & 34.45 $\pm$ 4.34 & 14.14  $\pm$ 1.92  & 20.31 $\pm$ 2.66 &   14.00   $\pm$  1.93 &   16.31 $\pm$ 2.20 &  4.14 $\pm$  0.80     \\   
  1.25 & 20.30 $\pm$ 1.30 & 10.28  $\pm$ 0.87  & 10.02 $\pm$ 0.86 &   7.16    $\pm$  0.78 &    7.43 $\pm$ 0.80 &  5.70 $\pm$  0.72     \\    
  1.55 & 14.62 $\pm$ 1.67 & 8.58   $\pm$ 1.12  & 6.04  $\pm$ 0.88 &   5.94    $\pm$  0.90 &    5.30 $\pm$ 0.85 &  3.37 $\pm$  0.67     \\   
  1.85 & 11.52 $\pm$ 0.62 & 6.65   $\pm$ 0.67  & 4.87  $\pm$ 0.62 &   4.21    $\pm$  0.66 &    3.32 $\pm$ 0.63 &  3.99 $\pm$  0.65     \\   
  2.25 & 8.16  $\pm$ 0.78 & 6.70   $\pm$ 0.77  & 1.46  $\pm$ 0.44 &   3.14    $\pm$  0.59 &    2.93 $\pm$ 0.58 &  2.09 $\pm$  0.52     \\   
  2.75 & 5.30  $\pm$ 1.07 & 3.78   $\pm$ 0.86  & 1.51  $\pm$ 0.50 &   2.27    $\pm$  0.64 &    1.01 $\pm$ 0.45 &  2.02 $\pm$  0.60     \\  
\end{tabular}
\end{table*}
\end{center}
\renewcommand{\arraystretch}{1}

\section{Discussion}
\label{sec:discussion}

The evidence collected in the previous section suggests that there is a strong evolution
in the morphological properties (both quantitative and qualitative) of the massive galaxy
population over time. At high redshift, in agreement with the theoretical expectation, the dominant
morphological classes of galaxies are late-types and peculiars. Consequently, the morphological
and structural types of the majority of the massive galaxies at a given epoch has dramatically evolved
as cosmic time increases. 

Two effects could play a role towards explaining this significant change
of the dominant morphological class. On one hand, the galaxies that are progressively been
added into the family of massive objects (i.e. by the merging of less massive galaxies)
are incorporated with already spheroidal morphologies. On the other hand, the already
old massive galaxies can also evolve
towards spheroidal morphologies due to frequent mergers. For instance, frequent minor
mergers (L\'opez-Sanjuan et al. 2010, Kaviraj 2010, L\'opez-Sanjuan et al. 2011, Bluck et al. 2012)  with the
massive galaxy population may destroy existing stellar disks, and also would be
responsible for the appearance of long tails in their luminosity profiles. This scenario
could explain why the evolution towards spheroid-like morphologies is stronger when we use
the S\'ersic index $n$ instead of the visual classification. 

In fact, the surface
brightness of nearby massive ellipticals are well described with large S\'ersic indices
due to their bright outer envelopes. These wings, however, seem to disappear at higher
and higher redshifts (see Table \ref{tab:structural_data} or Figure \ref{fig:mean_vs_median}) just leaving the inner (core) region of the massive galaxies
(Bezanson et al. 2009; Hopkins et al. 2009; van Dokkum et al. 2010; Carrasco, Conselice \&
Trujillo 2010). The disappearance of these outer envelopes is also connected with the dramatic size
evolution reported in previous works (see e.g. Trujillo et al. 2007; Buitrago et al. 2008; Van Dokkum et al. 2010; Trujillo, Ferreras \& De la Rosa 2011; McLure et al. 2012). Consequently, it is
not only that the typical morphology of the massive galaxy population is changing with
redshift, but also that there is a progressive build-up of their outer envelopes, making the
morphological evolution appears more dramatic when we use the S\'ersic index instead of
the visual classification as a morphological segregator.

An open question is whether we are also witnessing the progressive development of bulges with redshift. There are many indications which tell us this evolution is taking place. For instance Azzollini, Beckman \& Trujillo (2009) investigated the luminosity profiles of massive ($M_{\ast} > 10^{10} M_{\odot}$ in this case) disks at $0 < z < 1$. Some of these galaxies clearly show an excess of light in their surface brightness profiles in the reddest band which was not present for the bluest. Complementary, Dom\'inguez-Palmero \& Balcells (2008) reported nuclear star formation must lead to bulge growth within disks for HST ACS observations in a similar redshift range. However, we noticed there is a lack of studies focussing on the bulge growth with redshift, and its impact in the observed galaxy S\'ersic index.

If we were just using the information contained in the change of the fraction of
morphological types with redshift we would be tempted to explain the morphological
evolution as being just a consequence of a transformation from one class to another,
however, the evolution in the number density of all the classes suggests a more complex scenario. 
In fact, one of the results we can conclude from the evolution of the number
densities of all the massive galaxy classes is that high-z massive disk-like galaxies cannot be
the only progenitors of present-day massive spheroid-like galaxies. They are just simply
not enough in number to explain the large increase of the number density of elliptical
galaxies at low redshifts.

All the morphological classes, maybe with the exception of irregular/merging galaxies,
have increased their number densities with cosmic time. 
These irregular/merging galaxies also have shorter structural life-spans than the other
types, and therefore the galaxies in this class must be replenished through time.
This emergence of massive galaxies
is more efficient (by a factor $\sim$2) at creating spheroid-like galaxies than disk-like
objects from z$\sim$1 to now (see Fig. \ref{fig:percentages}). The reason why the formation of elliptical galaxies is more
efficient at recent times than it was in the past is possibly linked to the
lower availability of gas during mergers creating new galaxies (Khochfar
\& Silk 2006, 2009, Shankar et al. 2010, Eliche-Moral et al. 2010), and thus more
likely to contain denser and more concentrated light profiles (Bluck et al. 2012). 

It has been reported lately that, apart from the increasing number density of massive galaxies with redshift, the number density of quiescent galaxies among them has risen by a factor of five or even more (Brammer et al. 2011, Bell et al. 2011, Cassata et al. 2011, Barro et al. 2012). There is evidence about this quiescence is correlated with the S\'ersic index at all redshifts (Bell et al. 2011). This is in perfect agreement with our results, as we undoubtedly observe an increment of the S\'ersic index amid our sample towards low z. Moreover, it is reasonable to expect that the level of star formation must have an impact not only in the Sersic index but also in the galaxy morphology. Following the argument in favour of the appearance of massive and passive galaxies with cosmic time, it is nonetheless difficult to explain why the number densities of late-type massive galaxies slightly rise with redshift. However, we could reconcile the apparently contradictory facts by advocating the large number of passive disks found at high z (McGrath et al. 2008, Messias et al. in preparation). As stated in McLure et al. (2012), this reflects that different mechanisms may account for the quenching of star formation and the morphological transformation of massive galaxies.

If the increment of the S\'ersic index with time/redshift (and thus the prominence of bulge and outer envelopes) is linked with the availability of minor companions that will eventually merge with the main galaxy, then spheroid-like massive galaxies must have on average more of them, as it has been observationally confirmed (M\'armol-Queralt\'o et al. 2012). This satellite infalling would explain why these early-type/spheroid-like massive galaxies are not completely devoid of star formation (P\'erez-Gonz\'alez et al. 2008b, Cava et al. 2010, Bauer et al. 2011, Viero et al. 2012). The risen of massive early type galaxies appear to be a natural consequence of galaxy mass assembly. The contribution of the various physical mechanisms on changing the galaxy morphology and the existence of a considerable fraction of massive late type galaxies at low z still challenge our vision of the nature of massive galaxies. 

\section{Summary}
\label{summary}

Using a large compilation of massive ($M\geq10^{11} h_{70}^{-2} M_{\odot}$) galaxies
($\sim$1100 objects) since z$\sim$3 we have addressed the issue of the morphological
change of this population with time. We have found that there is a
profound transformation in the morphological content of massive galaxies during this
cosmic interval. Massive galaxies were typically disk-like in shape at z$\gtrsim$1 and
early type galaxies have been only the predominant massive class since that epoch. The
fraction of early-type morphologies in massive galaxies has changed from $\sim$20-30\% at z$\sim$3 to $\sim$70\%
at z=0 (see Figure \ref{fig:percentages}).

We have addressed the morphological transformation of the massive galaxies using a
quantitative approach, based on S\'ersic fits to the surface brightness distribution of the galaxies,
and a qualitative approach based on visual classifications. Both analyses agree on a clear
morphological change in the dominant morphological class with time. In particular, the
quantitative approach, which uses the S\'ersic index as a morphological segregator, shows
that the number of galaxies with low S\'ersic index at high-z was higher than in the
present day universe. We interpret this as a consequence of two phenomena: a decrease in the
number of early-type galaxies at higher redshift plus an intrinsic decrease of the
S\'ersic index values of those massive galaxies at earlier cosmic times due to the loss
of their extended envelopes.

\section{Acknowledgements}
\label{sec:ack}

This paper would have not been possible without hospitality of the IAC, in particular that of Ignacio Trujillo and
Jes\'{u}s Falc\'{o}n-Barroso. We gratefully thank the contribution of the anonymous referee
who helped our scientific discussion, and our editor Keith T. Smith who promptly answered all our requests. We acknowledge the help of Jes\'{u}s Varela on
finding the necessary information of the SDSS sample, Judit Bakos for assistance with SDSS
imaging and sky substraction and also Juan E. Betancort and In\'es Flores-Cacho 
for their scientific feedback, apart from
the contributions from the GNS team. We gratefully thank Steven Bamford, Nathan Bourne,
Carlos L\'{o}pez-Sanjuan and Pablo P\'erez-Gonz\'alez for useful discussions. FB acknowledges the support of the European Research Council. This work
has been supported by the STFC, NASA through NASA/STScI grant HST-GO11082,  the
Leverhulme Trust, and the Programa Nacional de Astronom\'ia y Astrof\'isica of the Spanish
Ministry of Science and Innovation under grant AYA2010-21322-C03-02.

\newpage 

\section*{Appendix A: GOODS NICMOS Survey massive galaxies simulations}
\label{sec:appendix}

The purpose of this Appendix is to explore the robustness of the structural parameters of
the massive galaxies (1.7$<$z$<$3) in our GNS sample (H-band, F160W filter, HST NICMOS-3
camera, 3 orbits depth; Conselice et al. 2011). As explained in the main text of the
paper, a set of simulations similar to the ones presented here were already conducted for
the ACS imaging used to analyze the galaxies in the redshift interval 0.2$<$z$<$2
(Trujillo et al. 2007). To identity the ranges of the structural parameters to explore in
our simulations, we use as a guide the ranges found in the quantitative morphological
analysis based on GALFIT of the real GNS massive galaxies (Buitrago et al. 2008). These
were:
$$0.15   < r_{e} (arcsec) < 0.61 \nonumber$$ 
$$0.34   < n              < 4.62 \nonumber$$
$$0.19   < ar             < 0.92 \nonumber$$
$$-84.03 < pa             < 85.28 \nonumber$$
$$20.5   < H_{AB}(mag)    < 24 \nonumber$$ 
\noindent where $r_{e}$, $n$, $ar$, $pa$, $H_{AB}$ stand for effective radius, S\'{e}rsic
index, axis ratio, position angle and derived $H_{AB}$-band magnitude. The only exceptions
were one galaxy with $n=0.17$ and two others with $24 < H_{AB} < 24.5$. Taking these
quantities into account, we simulated 16000 galaxies utilising the IDL routines built for
H\"au\ss ler et al. (2007), with the structural parameters randomized within these ranges:
$$0.15   < r_{e} (arcsec) < 2.0 \nonumber$$
$$0.25   < n              < 8.0 \nonumber$$
$$0.1    < ar             < 1.0 \nonumber$$
$$-90    < pa             < 90  \nonumber$$
$$20     < H_{AB}(mag)    < 25 \nonumber$$ 
The structural parameters of the mock galaxies were distributed linearly along the full
parameter space, except for the effective radii which were logarithmically sampled as we
specially wanted to explore objects with small angular radii due to the observed compactness
of massive galaxies at high-z.

Images of every mock galaxy were created placing these objects randomly in the GNS
pointings. We placed a mock galaxy on each GNS pointing for every simulation in order to avoid altering
the typical density (i.e. the number of neighbour galaxies) of the GNS imaging. Each model
galaxy (i.e. the 2D surface brightness distribution following the S\'ersic function) was
convolved with a representative PSF. Specifically we used one of the five natural stars
which were utilised in Buitrago et al. (2008). To obtain errors in the same way as in that
paper, we also ran GALFIT using these five different stars and then taking the mean
values. 

\begin{figure*}
\vspace{1cm}
\rotatebox{0}{
\includegraphics[angle=0,width=1.\linewidth]{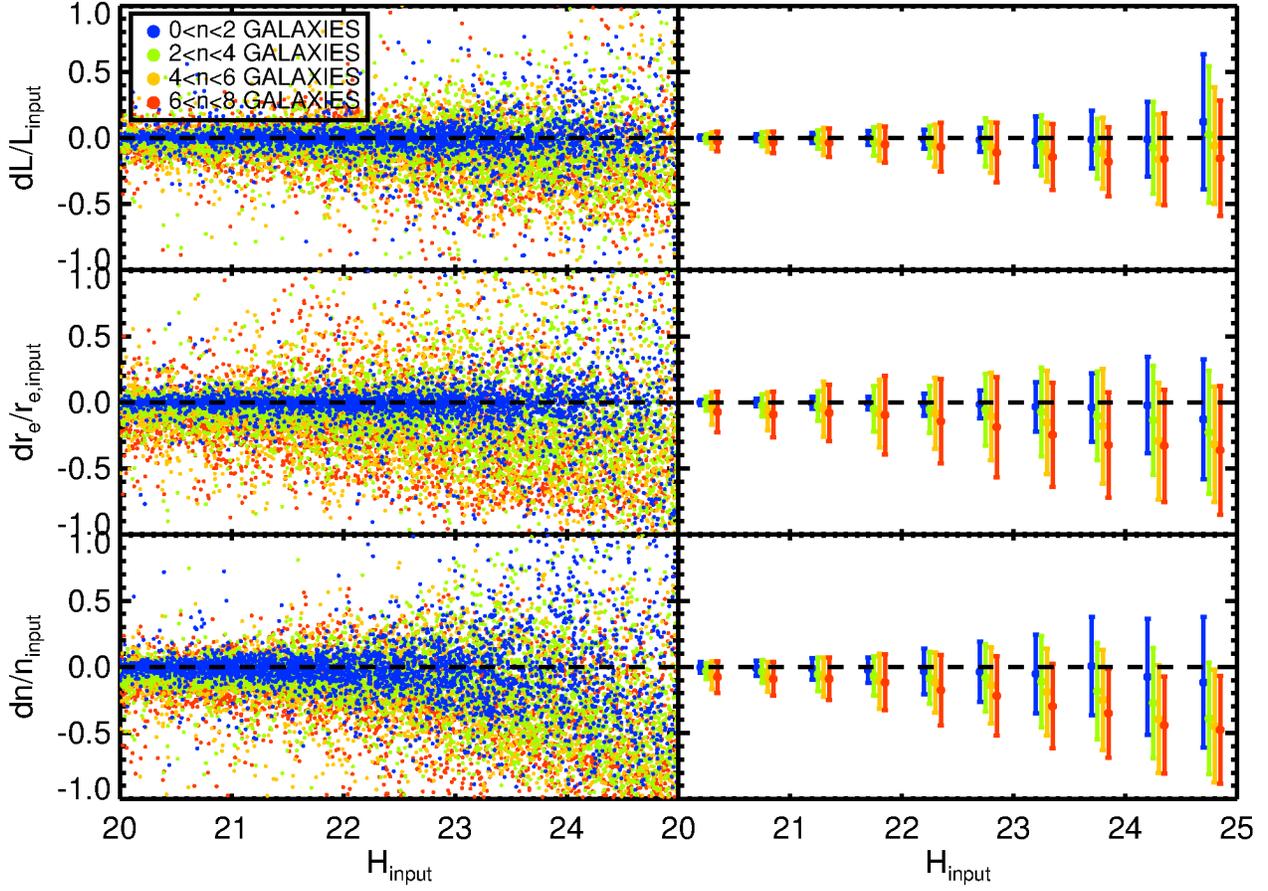}}
\vspace{-0.5cm}

\caption{Relative errors - \textit{(output-input)/input} - of the structural parameters (magnitude,
effective radius and S\'{e}rsic index) of our simulated GNS galaxies. The right column shows the means in
bins of 0.5 mag (with a $5\sigma$ outlier-resistant determination), being the error bars the standard
deviation of the sample. The information in this plot is tabulated in Table \ref{tab:str_param_mag_best_psf}.}
\label{fig:simus_mag_best_psf}

\end{figure*}

\begin{figure*}
\rotatebox{0}{
\includegraphics[angle=0,width=1.\linewidth]{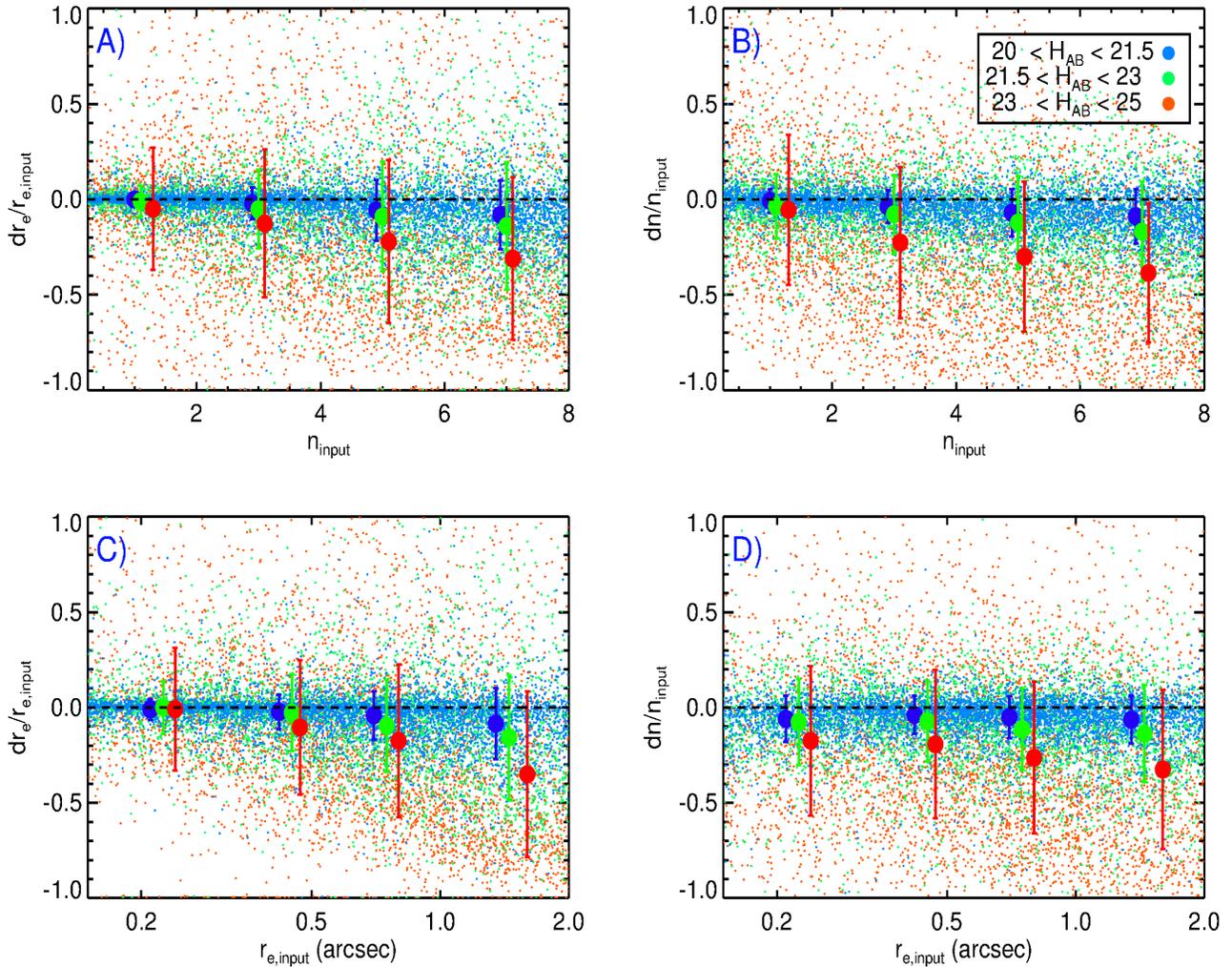}}

\caption{Relative errors - \textit{output-input/input} - of the effective radius (first column) and the
S\'{e}rsic index (second column) as a function of the input S\'{e}rsic index (first row) and the input effective
radius (second row). Galaxies are coloured according to their magnitude. For the sake of clarity, mean values
(derived with a $5\sigma$ outlier-resistant determination) where added using 4 intervals in effective radius
and S\'{e}rsic index, with the error bars being their standard deviation. S\'{e}rsic index intervals are $0 <
n < 2$, $2 < n < 4$, $4 < n < 6$ and $6 < n < 8$. Effective radius intervals are $0.15" < r_{e} < 0.3"$,
$0.3" < r_{e} < 0.6"$, $0.6" < r_{e} < 0.9"$ and $0.9" < r_{e} < 2"$. Note that the colour of these mean
points is the same as the one of the galaxy individual points. The information in this plot is tabulated in Table \ref{tab:str_param_n_and_re_best_psf}.}
\label{fig:simus_re_and_n_best_psf}

\end{figure*}

\begin{figure*}
\vspace{1cm}
\rotatebox{0}{
\includegraphics[angle=0,width=1.\linewidth]{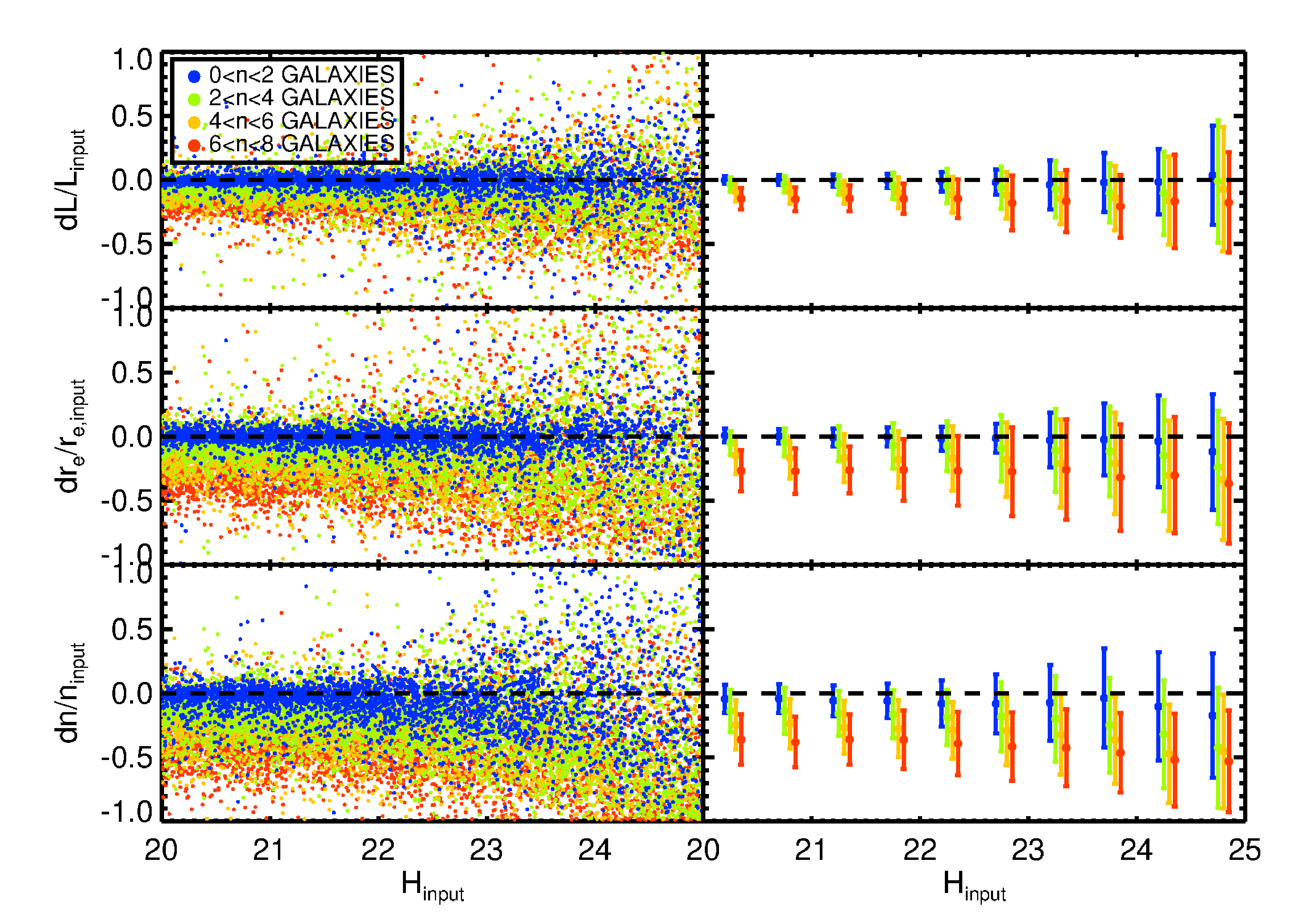}}
\vspace{-0.5cm}

\caption{Same as in Figure \ref{fig:simus_mag_best_psf}, but using this time as the output parameters the mean values of the fits
retrieved based on 5 different natural stars as PSFs. The results of this figure are tabulated in Table \ref{tab:str_param_mag}.}
\label{fig:simus_mag}

\end{figure*}

\begin{figure*}
\rotatebox{0}{
\includegraphics[angle=0,width=1.\linewidth]{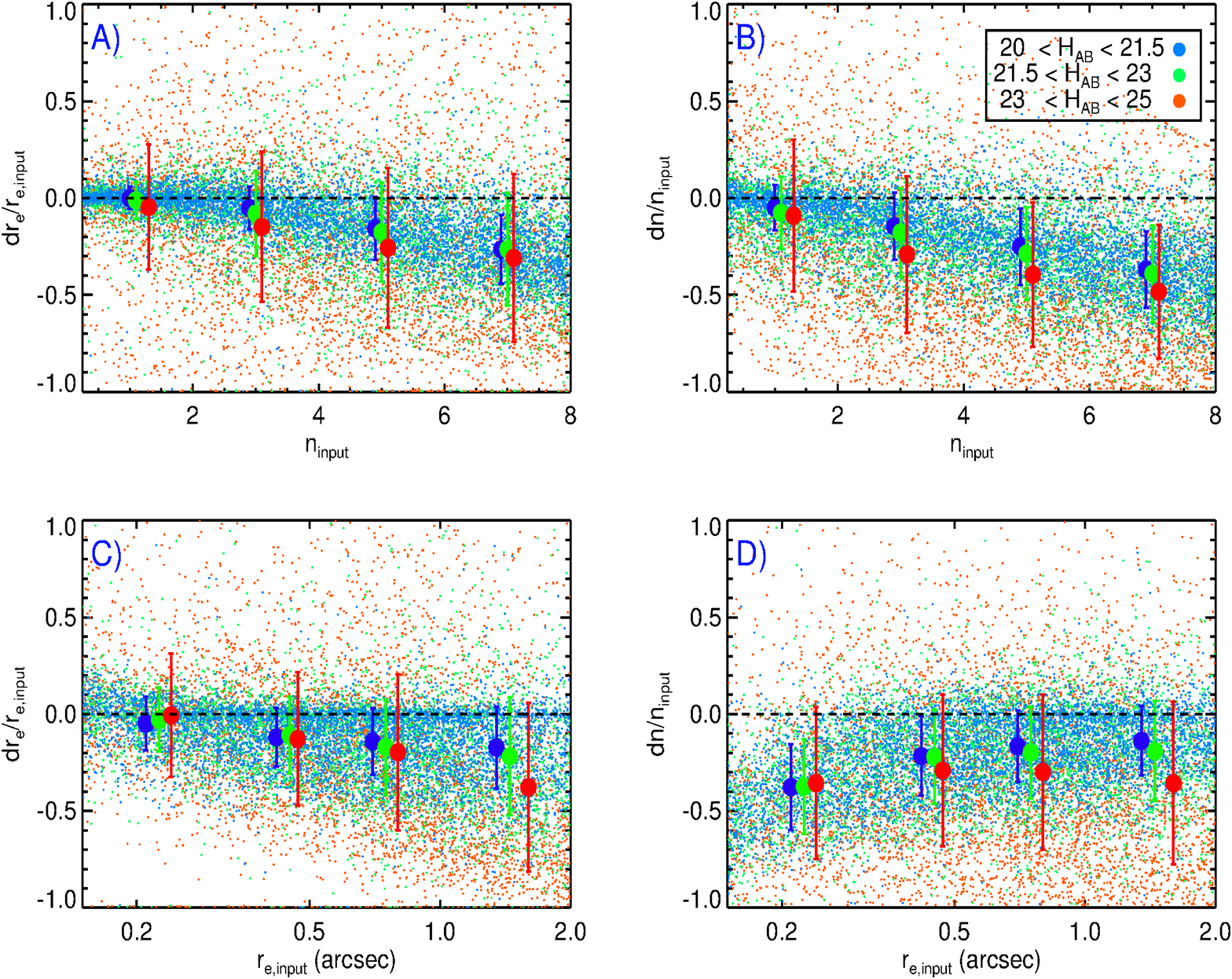}}

\caption{Same as Figure \ref{fig:simus_re_and_n_best_psf}, but using this time as the output parameters the mean values of the fits
retrieved based on 5 different natural stars as PSFs. The results of this figure are tabulated in Table \ref{tab:str_param_n_and_re}.}
\label{fig:simus_re_and_n}

\end{figure*}

We have identified that the main source of uncertainty in the NICMOS data is the change of
the PSF among the different pointings. To illustrate how this affects the
recovery of the structural parameters we first present in Figure \ref{fig:simus_mag_best_psf} and \ref{fig:simus_re_and_n_best_psf} how our
parameters are recovered when we use the same PSF for creating and recovering the mock
galaxies. In Figure \ref{fig:simus_mag} and \ref{fig:simus_re_and_n}, we show the effects on the parameters when we compare
the input values with the average values obtained using the five different PSFs. 

In Figures \ref{fig:simus_mag_best_psf} and \ref{fig:simus_mag} we show the relationship between the relative errors in the structural
parameters (magnitude, effective radius and S\'{e}rsic index) versus the galaxy input
magnitude. The relative errors are calculated as \textit{(output-input)/input} (i. e.,
a negative \% refer to cases where the output is smaller than the input and vice-versa). The
left column of the plot displays the structural parameters of individual mock galaxies, whereas
the right column shows their means in bins of 0.5 mag. The mean values of the structural
parameters were derived using a robust method which removes the $5\sigma$ outliers. Error
bars represent the standard deviation of the sample. To appreciate how the effect on the
structural parameter is linked to the input S\'ersic index of the mock galaxies we split
the sample into four groups  ($0 < n < 2$, $2 < n < 4$, $4 < n < 6$ and $6 < n < 8$). The
results shown in Fig. \ref{fig:simus_mag} are tabulated in the Table \ref{tab:str_param_mag}. At increasing S\'ersic index values, the
recovery of the structural parameters is more affected. We note that galaxies with low S\'ersic index
are well recovered down to our faintest magnitude. An average galaxy in our GNS sample (H=22.5 mag) with an
input S\'ersic index of n=4 will have its effective radius biased only by a $\sim$10\% and its S\'ersic index
around $\sim$20\%. 

In addition to the dependence of the apparent magnitude of the objects on recovering their structural parameters,
in Figures \ref{fig:simus_re_and_n_best_psf} and \ref{fig:simus_re_and_n} we explore what is the effect of the size (lower row) and intrinsic shape (upper row) for this
matter. Galaxies are colour coded in these figures according to their magnitude. Combining the information contained in Fig. \ref{fig:simus_mag}
and \ref{fig:simus_re_and_n}, we find that the key parameters for retrieving accurate structural parameters are the apparent
magnitude and the S\'ersic index. The effective radius of the objects play a minor role. The results show on
Fig. \ref{fig:simus_re_and_n} are tabulated in the Table \ref{tab:str_param_n_and_re}.


The most interested output of our tests using the mean value from the fits of different PSFs is that the size
of the source plays now a fundamental role at characterising the error on the structural parameters. As
expected, large sources are less affected by changes in the PSF and the bias on the structural
parameters remain basically the same as when we use just a single PSFs. However, at smaller sizes the
effect of not knowing accurately the PSF at the source implies that the S\'{e}rsic index uncertainty is large,
although sizes are retrieved accurately. Summarising, neither any effect or combination of effects is
 large enough to modify the main results of this paper. 
Moreover, as
stated on the main text of this paper, we use these simulations to correct, based on the observed (output)
apparent magnitude, effective radius and S\'ersic index, the structural parameters presented in this work.


\renewcommand{\arraystretch}{1.2}
\begin{center}
\begin{table*}
\caption{Relative errors (\%) and standard deviations on the structural parameters depending on the apparent magnitude (see Fig. \ref{fig:simus_mag_best_psf}).}
\label{tab:str_param_mag_best_psf}
\hspace{0.4in}
\begin{tabular}{c|c@{}|c@{}|c@{}|c@{}}
\hline
 & $0 < n < 2$ & $2 < n < 4$ & $4 < n < 6$ & $6 < n < 8$ \\
\hline
\hline
$20.0 < H_{AB,input}(mag) < 20.5$\\
\hline
$\delta L/L $ & $  0.39\pm  1.75$ & $ -0.53\pm  3.21$ & $ -1.85\pm  6.24$ & $ -2.62\pm  7.36$\\
$\delta r_{e}/r_{e}$ & $  0.09\pm  2.03$ & $ -1.16\pm  5.27$ & $ -4.63\pm 12.35$ & $ -7.19\pm 15.54$\\
$\delta n/n$ & $ -0.34\pm  4.08$ & $ -3.22\pm  6.20$ & $ -6.25\pm 10.59$ & $ -7.62\pm 12.07$\\
\hline
$20.5 < H_{AB,input}(mag) < 21.0$\\
\hline
$\delta L/L $ & $  0.21\pm  2.61$ & $ -0.99\pm  5.46$ & $ -2.95\pm  7.61$ & $ -3.43\pm  8.20$\\
$\delta r_{e}/r_{e}$ & $  0.01\pm  2.98$ & $ -2.33\pm  8.31$ & $ -6.78\pm 14.35$ & $ -9.16\pm 17.30$\\
$\delta n/n$ & $ -0.97\pm  5.63$ & $ -3.51\pm  8.48$ & $ -7.69\pm 11.25$ & $ -9.13\pm 12.68$\\
\hline
$21.0 < H_{AB,input}(mag) < 21.5$\\
\hline
$\delta L/L $ & $ -0.12\pm  3.85$ & $ -1.34\pm  6.93$ & $ -2.60\pm 10.14$ & $ -3.62\pm 10.74$\\
$\delta r_{e}/r_{e}$ & $ -0.40\pm  4.58$ & $ -2.81\pm 11.39$ & $ -5.08\pm 20.67$ & $ -7.85\pm 21.32$\\
$\delta n/n$ & $ -1.42\pm  8.21$ & $ -5.65\pm 12.94$ & $ -7.72\pm 15.04$ & $ -9.08\pm 16.12$\\
\hline
$21.5 < H_{AB,input}(mag) < 22.0$\\
\hline
$\delta L/L $ & $ -0.05\pm  5.17$ & $ -2.56\pm 10.93$ & $ -3.76\pm 13.39$ & $ -4.95\pm 13.55$\\
$\delta r_{e}/r_{e}$ & $ -0.50\pm  4.79$ & $ -4.98\pm 17.21$ & $ -8.46\pm 25.80$ & $ -9.54\pm 29.76$\\
$\delta n/n$ & $ -2.47\pm  9.68$ & $ -6.85\pm 14.97$ & $-10.83\pm 20.77$ & $-11.80\pm 21.14$\\
\hline
$22.0 < H_{AB,input}(mag) < 22.5$\\
\hline
$\delta L/L $ & $ -1.03\pm  7.28$ & $ -2.87\pm 12.06$ & $ -5.83\pm 15.87$ & $ -6.79\pm 18.50$\\
$\delta r_{e}/r_{e}$ & $ -1.70\pm  8.48$ & $ -5.33\pm 17.56$ & $ -8.22\pm 27.02$ & $-14.28\pm 32.00$\\
$\delta n/n$ & $ -3.55\pm 17.15$ & $ -8.50\pm 19.65$ & $-11.86\pm 22.91$ & $-17.65\pm 26.70$\\
\hline
$22.5 < H_{AB,input}(mag) < 23.0$\\
\hline
$\delta L/L $ & $ -1.46\pm  9.27$ & $ -3.40\pm 18.16$ & $ -6.97\pm 19.19$ & $-11.04\pm 22.58$\\
$\delta r_{e}/r_{e}$ & $ -1.48\pm 10.75$ & $ -5.55\pm 26.98$ & $-10.61\pm 33.02$ & $-18.60\pm 37.86$\\
$\delta n/n$ & $ -3.74\pm 22.80$ & $ -8.20\pm 25.31$ & $-13.99\pm 28.88$ & $-21.90\pm 29.83$\\
\hline
$23.0 < H_{AB,input}(mag) < 23.5$\\
\hline
$\delta L/L $ & $ -2.70\pm 18.92$ & $ -5.54\pm 22.91$ & $-10.57\pm 22.24$ & $-14.38\pm 25.00$\\
$\delta r_{e}/r_{e}$ & $ -3.38\pm 18.65$ & $ -7.10\pm 33.57$ & $-15.11\pm 38.94$ & $-24.55\pm 39.17$\\
$\delta n/n$ & $ -5.29\pm 29.66$ & $-11.14\pm 34.34$ & $-18.92\pm 32.87$ & $-29.79\pm 31.89$\\
\hline
$23.5 < H_{AB,input}(mag) < 24.0$\\
\hline
$\delta L/L $ & $ -1.28\pm 21.90$ & $ -8.12\pm 22.41$ & $-11.57\pm 26.88$ & $-17.93\pm 26.23$\\
$\delta r_{e}/r_{e}$ & $ -3.81\pm 26.00$ & $-10.21\pm 33.56$ & $-17.86\pm 43.47$ & $-32.17\pm 39.63$\\
$\delta n/n$ & $  0.61\pm 37.18$ & $-18.35\pm 36.38$ & $-24.90\pm 37.94$ & $-35.02\pm 33.59$\\
\hline
$24.0 < H_{AB,input}(mag) < 24.5$\\
\hline
$\delta L/L $ & $ -0.99\pm 28.30$ & $ -7.27\pm 34.80$ & $-16.06\pm 33.78$ & $-15.98\pm 34.87$\\
$\delta r_{e}/r_{e}$ & $ -2.13\pm 36.39$ & $-13.48\pm 41.48$ & $-29.16\pm 44.22$ & $-32.70\pm 42.43$\\
$\delta n/n$ & $ -7.57\pm 43.98$ & $-27.35\pm 41.82$ & $-39.53\pm 40.77$ & $-44.10\pm 36.75$\\
\hline
$24.5 < H_{AB,input}(mag) < 25.0$\\
\hline
$\delta L/L $ & $ 12.20\pm 51.04$ & $  2.50\pm 51.58$ & $ -5.73\pm 44.09$ & $-15.50\pm 43.68$\\
$\delta r_{e}/r_{e}$ & $-12.85\pm 45.18$ & $-22.74\pm 46.45$ & $-31.77\pm 43.55$ & $-36.15\pm 48.64$\\
$\delta n/n$ & $-11.73\pm 49.37$ & $-38.99\pm 42.35$ & $-44.67\pm 42.69$ & $-47.58\pm 40.68$\\
\hline
\end{tabular}
\end{table*}
\end{center}
\renewcommand{\arraystretch}{1}

\renewcommand{\arraystretch}{1.2}
\begin{center}
\begin{table*}
\caption{Relative errors (\%) and standard deviations on the structural parameters (see Fig. \ref{fig:simus_re_and_n_best_psf})}
\label{tab:str_param_n_and_re_best_psf}
\hspace{0.4in}
\begin{tabular}{c|c|c|c}
\hline
$\delta r_{e}/r_{e}$ & $20 < H_{AB}(mag) < 21.5$ & $21.5 < H_{AB}(mag) < 23$ & $23 < H_{AB}(mag) < 25$\\
\hline
$ 0 < n < 2 $  & $ -0.04\pm  3.09$ & $ -1.00\pm  7.97$ & $ -4.89\pm 31.91$\\
$ 2 < n < 4 $  & $ -2.14\pm  8.27$ & $ -5.13\pm 20.61$ & $-12.66\pm 38.79$\\
$ 4 < n < 6 $  & $ -5.71\pm 15.87$ & $ -9.08\pm 28.73$ & $-22.12\pm 42.84$\\
$ 6 < n < 8 $  & $ -8.16\pm 18.05$ & $-13.98\pm 33.45$ & $-31.04\pm 42.55$\\
\hline
$ 0.15 < r_{e}(arcsec) < 0.3 $& $ -0.94\pm  5.45$ & $  0.05\pm 13.79$ & $ -0.87\pm 32.27$\\
$ 0.3  < r_{e}(arcsec) < 0.6 $& $ -2.28\pm  9.02$ & $ -2.71\pm 20.11$ & $-10.47\pm 35.21$\\
$ 0.6  < r_{e}(arcsec) < 0.9 $& $ -4.15\pm 12.67$ & $ -9.55\pm 24.26$ & $-17.50\pm 40.12$\\
$ 0.9  < r_{e}(arcsec) < 2.0 $& $ -8.44\pm 18.68$ & $-15.66\pm 32.69$ & $-35.02\pm 43.53$\\
\hline
\hline
$\delta n/n$ & $20 < H_{AB} < 21.5$ & $21.5 < H_{AB} < 23$ & $23 < H_{AB} < 25$\\
\hline
$ 0 < n < 2 $  & $ -0.84\pm  6.23$ & $ -3.57\pm 16.86$ & $ -5.51\pm 39.42$\\
$ 2 < n < 4 $  & $ -3.75\pm  8.71$ & $ -7.83\pm 20.63$ & $-22.65\pm 39.74$\\
$ 4 < n < 6 $  & $ -7.08\pm 12.33$ & $-12.19\pm 24.38$ & $-30.01\pm 39.47$\\
$ 6 < n < 8 $  & $ -8.85\pm 14.25$ & $-16.94\pm 26.32$ & $-38.55\pm 36.30$\\
\hline
$ 0.15 < r_{e}(arcsec) < 0.3 $& $ -5.99\pm 12.04$ & $ -7.56\pm 22.54$ & $-17.43\pm 39.18$\\
$ 0.3  < r_{e}(arcsec) < 0.6 $& $ -3.93\pm 10.03$ & $ -7.20\pm 21.14$ & $-19.31\pm 38.89$\\
$ 0.6  < r_{e}(arcsec) < 0.9 $& $ -5.08\pm 11.05$ & $-11.31\pm 21.56$ & $-26.36\pm 39.79$\\
$ 0.9  < r_{e}(arcsec) < 2.0 $& $ -6.38\pm 12.51$ & $-13.74\pm 25.47$ & $-32.49\pm 41.78$\\
\hline
\end{tabular}
\end{table*}
\end{center}
\renewcommand{\arraystretch}{1}

\renewcommand{\arraystretch}{1.2}
\begin{center}
\begin{table*}
\caption{Relative errors (\%) and standard deviations on the structural parameters depending on the apparent magnitude using five
different PSFs (see Fig. \ref{fig:simus_mag}).}
\label{tab:str_param_mag}
\hspace{0.4in}
\begin{tabular}{c|c@{}|c@{}|c@{}|c@{}}
\hline
 & $0 < n < 2$ & $2 < n < 4$ & $4 < n < 6$ & $6 < n < 8$ \\
\hline
\hline
$20.0 < H_{AB,input}(mag) < 20.5$\\
\hline
$\delta L/L $ & $ -0.15\pm  2.95$ & $ -3.39\pm  5.48$ & $ -9.09\pm  7.90$ & $-14.50\pm  8.47$\\
$\delta r_{e}/r_{e}$ & $  0.79\pm  5.25$ & $ -4.95\pm  9.28$ & $-14.69\pm 14.79$ & $-26.56\pm 16.21$\\
$\delta n/n$ & $ -4.39\pm 10.97$ & $-13.84\pm 16.60$ & $-24.76\pm 19.20$ & $-36.22\pm 19.58$\\
\hline
$20.5 < H_{AB,input}(mag) < 21.0$\\
\hline
$\delta L/L $ & $ -0.14\pm  3.82$ & $ -3.44\pm  6.32$ & $ -9.31\pm  9.39$ & $-15.08\pm  9.23$\\
$\delta r_{e}/r_{e}$ & $  0.20\pm  5.36$ & $ -4.57\pm 10.89$ & $-16.51\pm 16.28$ & $-26.97\pm 17.87$\\
$\delta n/n$ & $ -4.16\pm 11.25$ & $-13.82\pm 17.91$ & $-23.84\pm 19.20$ & $-38.19\pm 19.64$\\
\hline
$21.0 < H_{AB,input}(mag) < 21.5$\\
\hline
$\delta L/L $ & $ -0.57\pm  4.92$ & $ -3.68\pm  7.55$ & $ -9.49\pm  9.25$ & $-14.29\pm 10.06$\\
$\delta r_{e}/r_{e}$ & $ -0.86\pm  7.00$ & $ -5.73\pm 13.66$ & $-16.66\pm 18.67$ & $-26.05\pm 18.35$\\
$\delta n/n$ & $ -5.76\pm 12.02$ & $-15.86\pm 17.54$ & $-26.63\pm 20.50$ & $-36.08\pm 19.52$\\
\hline
$21.5 < H_{AB,input}(mag) < 22.0$\\
\hline
$\delta L/L $ & $ -0.49\pm  5.53$ & $ -4.59\pm 10.36$ & $ -9.94\pm 12.10$ & $-14.68\pm 11.82$\\
$\delta r_{e}/r_{e}$ & $ -0.29\pm  7.33$ & $ -7.05\pm 17.72$ & $-17.91\pm 22.02$ & $-25.85\pm 24.09$\\
$\delta n/n$ & $ -6.04\pm 13.43$ & $-16.61\pm 18.69$ & $-27.84\pm 22.26$ & $-36.30\pm 22.71$\\
\hline
$22.0 < H_{AB,input}(mag) < 22.5$\\
\hline
$\delta L/L $ & $ -1.47\pm  7.60$ & $ -4.71\pm 13.41$ & $-11.55\pm 13.82$ & $-14.61\pm 15.32$\\
$\delta r_{e}/r_{e}$ & $ -1.75\pm  9.51$ & $ -7.38\pm 19.43$ & $-16.97\pm 25.13$ & $-26.68\pm 26.94$\\
$\delta n/n$ & $ -8.18\pm 17.99$ & $-18.82\pm 21.73$ & $-28.87\pm 22.24$ & $-39.18\pm 24.90$\\
\hline
$22.5 < H_{AB,input}(mag) < 23.0$\\
\hline
$\delta L/L $ & $ -1.67\pm 10.03$ & $ -5.89\pm 16.50$ & $-11.89\pm 18.38$ & $-17.96\pm 21.49$\\
$\delta r_{e}/r_{e}$ & $ -1.28\pm 11.40$ & $ -9.24\pm 26.02$ & $-17.99\pm 29.22$ & $-27.32\pm 34.57$\\
$\delta n/n$ & $ -8.25\pm 22.92$ & $-18.29\pm 26.96$ & $-29.25\pm 27.00$ & $-41.66\pm 26.64$\\
\hline
$23.0 < H_{AB,input}(mag) < 23.5$\\
\hline
$\delta L/L $ & $ -3.80\pm 19.13$ & $ -7.12\pm 22.24$ & $-14.63\pm 19.75$ & $-16.61\pm 24.12$\\
$\delta r_{e}/r_{e}$ & $ -2.84\pm 21.33$ & $-10.78\pm 32.29$ & $-20.79\pm 34.39$ & $-25.78\pm 39.19$\\
$\delta n/n$ & $ -7.49\pm 29.62$ & $-20.09\pm 33.28$ & $-32.47\pm 31.11$ & $-42.58\pm 29.90$\\
\hline
$23.5 < H_{AB,input}(mag) < 24.0$\\
\hline
$\delta L/L $ & $ -1.89\pm 22.85$ & $-10.19\pm 22.94$ & $-14.19\pm 25.32$ & $-20.52\pm 24.32$\\
$\delta r_{e}/r_{e}$ & $ -2.21\pm 27.87$ & $-11.57\pm 35.12$ & $-20.76\pm 39.58$ & $-31.83\pm 41.51$\\
$\delta n/n$ & $ -3.87\pm 38.56$ & $-24.95\pm 36.87$ & $-36.53\pm 34.02$ & $-46.46\pm 31.09$\\
\hline
$24.0 < H_{AB,input}(mag) < 24.5$\\
\hline
$\delta L/L $ & $ -1.50\pm 25.40$ & $-10.23\pm 32.27$ & $-16.10\pm 34.15$ & $-16.71\pm 36.33$\\
$\delta r_{e}/r_{e}$ & $ -3.60\pm 35.90$ & $-14.78\pm 43.22$ & $-30.37\pm 42.96$ & $-30.21\pm 45.39$\\
$\delta n/n$ & $-10.32\pm 42.23$ & $-32.03\pm 42.13$ & $-47.21\pm 38.40$ & $-51.98\pm 36.24$\\
\hline
$24.5 < H_{AB,input}(mag) < 25.0$\\
\hline
$\delta L/L $ & $  3.48\pm 38.67$ & $ -0.99\pm 47.79$ & $ -7.13\pm 48.32$ & $-17.63\pm 39.12$\\
$\delta r_{e}/r_{e}$ & $-11.83\pm 45.10$ & $-24.02\pm 44.12$ & $-33.58\pm 47.25$ & $-36.64\pm 47.00$\\
$\delta n/n$ & $-17.46\pm 48.50$ & $-42.41\pm 46.73$ & $-45.35\pm 44.40$ & $-53.14\pm 39.50$\\
\hline
\end{tabular}
\end{table*}
\end{center}
\renewcommand{\arraystretch}{1}

\renewcommand{\arraystretch}{1.2}
\begin{center}
\begin{table*}
\caption{Relative errors (\%) and standard deviations on the structural parameters using five
different PSFs (see Fig. \ref{fig:simus_re_and_n})}
\label{tab:str_param_n_and_re}
\hspace{0.4in}
\begin{tabular}{c|c|c|c}
\hline
$\delta r_{e}/r_{e}$ & $20 < H_{AB}(mag) < 21.5$ & $21.5 < H_{AB}(mag) < 23$ & $23 < H_{AB}(mag) < 25$\\
\hline
$ 0 < n < 2 $  & $  0.06\pm  5.77$ & $ -1.16\pm  9.59$ & $ -4.66\pm 32.39$\\
$ 2 < n < 4 $  & $ -5.12\pm 11.07$ & $ -8.03\pm 21.95$ & $-14.81\pm 38.86$\\
$ 4 < n < 6 $  & $-15.72\pm 16.22$ & $-17.63\pm 25.57$ & $-25.64\pm 41.12$\\
$ 6 < n < 8 $  & $-26.47\pm 17.73$ & $-26.40\pm 29.20$ & $-30.86\pm 43.40$\\
\hline
$ 0.15 < r_{e}(arcsec) < 0.3 $& $ -4.88\pm 13.91$ & $ -2.74\pm 16.53$ & $ -0.63\pm 31.95$\\
$ 0.3  < r_{e}(arcsec) < 0.6 $& $-12.01\pm 15.03$ & $-10.85\pm 19.95$ & $-12.85\pm 34.54$\\
$ 0.6  < r_{e}(arcsec) < 0.9 $& $-14.19\pm 17.17$ & $-16.98\pm 24.64$ & $-19.68\pm 40.32$\\
$ 0.9  < r_{e}(arcsec) < 2.0 $& $-17.15\pm 21.19$ & $-21.85\pm 30.48$ & $-37.75\pm 43.51$\\
\hline
\hline
$\delta n/n$ & $20 < H_{AB} < 21.5$ & $21.5 < H_{AB} < 23$ & $23 < H_{AB} < 25$\\
\hline
$ 0 < n < 2 $  & $ -4.89\pm 11.68$ & $ -7.60\pm 18.65$ & $ -9.07\pm 39.19$\\
$ 2 < n < 4 $  & $-14.57\pm 17.28$ & $-17.79\pm 22.82$ & $-29.09\pm 40.43$\\
$ 4 < n < 6 $  & $-25.06\pm 19.75$ & $-28.64\pm 23.92$ & $-39.53\pm 37.22$\\
$ 6 < n < 8 $  & $-36.86\pm 19.70$ & $-39.00\pm 24.87$ & $-48.32\pm 34.49$\\
\hline
$ 0.15 < r_{e}(arcsec) < 0.3 $& $-37.61\pm 22.17$ & $-37.41\pm 24.34$ & $-35.55\pm 39.30$\\
$ 0.3  < r_{e}(arcsec) < 0.6 $& $-21.64\pm 20.35$ & $-22.08\pm 24.19$ & $-29.12\pm 39.11$\\
$ 0.6  < r_{e}(arcsec) < 0.9 $& $-16.58\pm 18.55$ & $-19.83\pm 23.63$ & $-29.86\pm 40.01$\\
$ 0.9  < r_{e}(arcsec) < 2.0 $& $-13.81\pm 17.93$ & $-19.11\pm 25.80$ & $-35.62\pm 42.06$\\
\hline
\end{tabular}
\end{table*}
\end{center}
\renewcommand{\arraystretch}{1}


%
%
%
%
%

\begin{thebibliography}{99}
\bibitem[\protect\citeauthoryear{}{}]{b1} Abazajian K. N. et al. 2009, ApJS, 182, 543
\bibitem[\protect\citeauthoryear{}{}]{b1} Azzollini R., Beckman J. E. \& Trujillo I., A\&A, 501, 119
\bibitem[\protect\citeauthoryear{}{}]{b1} Baldry I.~K. et al. 2004, ApJ, 600, 681
\bibitem[\protect\citeauthoryear{}{}]{b1} Barden M. et al. 2005, ApJ, 635, 959
\bibitem[\protect\citeauthoryear{}{}]{b1} Barden M. et al. 2012, MNRAS, 422, 449
\bibitem[\protect\citeauthoryear{}{}]{b1} Barger, A. J., Cowie L. L. \& Wang W.-H. 2008, ApJ, 689, 687
\bibitem[\protect\citeauthoryear{}{}]{b1} Barro G. et al. 2011, ApJS, 193, 13
\bibitem[\protect\citeauthoryear{}{}]{b1} Barro G. et al. 2012, arXiv: 1206.5000
\bibitem[\protect\citeauthoryear{}{}]{b1} Bauer A. E. et al. 2011, MNRAS, 417, 289
\bibitem[\protect\citeauthoryear{}{}]{b1} Bell E. F. et al. 2012, ApJ, 753, 167
\bibitem[\protect\citeauthoryear{}{}]{b1} Ben\'itez N., 2000, ApJ, 536, 571
\bibitem[\protect\citeauthoryear{}{}]{b1} Bertin E. \& Arnouts S., 1996, A\&AS, 117, 393
\bibitem[\protect\citeauthoryear{}{}]{b1} Bezanson R. et al., ApJ, 697, 1290
\bibitem[\protect\citeauthoryear{}{}]{b1} Blanton M. R. et al. 2005, ApJ, 629, 143
\bibitem[\protect\citeauthoryear{}{}]{b1} Blanton M: R. \& Roweis S. 2007, AJ, 133, 734
\bibitem[\protect\citeauthoryear{}{}]{b1} Bluck A. F. L., Conselice C. J., Bouwens R. J., Daddi E., Dickinson M., Papovich C., Haojing Y., 2009, MNRAS, 394, l51
\bibitem[\protect\citeauthoryear{}{}]{b1} Bluck A. F. L. et al. 2012, ApJ, 747, 34
\bibitem[\protect\citeauthoryear{}{}]{b1} Bournaud F. et al. 2011, ApJ, 730, 4
\bibitem[\protect\citeauthoryear{}{}]{b1} Brammer, Gabriel B.; van Dokkum, Pieter G.; Coppi, Paolo, 2008, ApJ, 686, 1503
\bibitem[\protect\citeauthoryear{}{}]{b1} Brammer G. B. et al., 2011, ApJ, 739, 24
\bibitem[\protect\citeauthoryear{}{}]{b1} Bruzual G. \& Charlot S., 2003, MNRAS, 344, 1000
\bibitem[\protect\citeauthoryear{}{}]{b1} Buitrago F. et al. 2008, ApJ, 687, L61
\bibitem[\protect\citeauthoryear{}{}]{b1} Bundy K., Ellis R. S. \& Conselice C. J., 2005, ApJ, 625, 621
\bibitem[\protect\citeauthoryear{}{}]{b1} Bundy K. et al., 2006, ApJ, 651, 120
\bibitem[\protect\citeauthoryear{}{}]{b1} Cameron E. et al. 2011, ApJ, 743, 146
\bibitem[\protect\citeauthoryear{}{}]{b1} Carrasco E. R., Conselice C. J. \& Trujillo I. 2010, MNRAS, 405, 2253
\bibitem[\protect\citeauthoryear{}{}]{b1} Cassata P. et al. 2011, ApJ, 743, 96
\bibitem[\protect\citeauthoryear{}{}]{b1} Cava A. et al. 2010, MNRAS, 409, L19
\bibitem[\protect\citeauthoryear{}{}]{b1} Chabrier G., 2003, PASP, 115, 763
\bibitem[\protect\citeauthoryear{}{}]{b1} Cimatti A. et al. 2008, A\&A, 482, 21
\bibitem[\protect\citeauthoryear{}{}]{b1} Cirasuolo M., McLure R. J., Dunlop J. S., Almaini O., Foucaud S., Simpson C., 2010, MNRAS, 401, 1166
\bibitem[\protect\citeauthoryear{}{}]{b1} Cole S. et al. 2001, MNRAS, 326, 255
\bibitem[\protect\citeauthoryear{}{}]{b1} Collister A. A. \& Lahav O., 2004, PASP, 116, 345
\bibitem[\protect\citeauthoryear{}{}]{b1} Conselice C. J., 2006, MNRAS, 373, 1389
\bibitem[\protect\citeauthoryear{}{}]{b1} Conselice C. J. et al., 2007, MNRAS, 381, 962
\bibitem[\protect\citeauthoryear{}{}]{b1} Conselice C. J., Bundy K., U V., Eisenhardt P., Lotz J., Newman, J., 2008, MNRAS, 383, 1366
\bibitem[\protect\citeauthoryear{}{}]{b1} Conselice C. J., Yang C., Bluck A. F. L., 2009, MNRAS, 394, 1956
\bibitem[\protect\citeauthoryear{}{}]{b1} Conselice C. J. et al. 2011a, MNRAS, 413, 80
\bibitem[\protect\citeauthoryear{}{}]{b1} Daddi E., et al. 2007, ApJ, 670, 156
\bibitem[\protect\citeauthoryear{}{}]{b1} Davis M. et al. 2003, SPIE, 4834, 161
\bibitem[\protect\citeauthoryear{}{}]{b1} Dekel A. et al. 2009, Nature, 457, 451
\bibitem[\protect\citeauthoryear{}{}]{b1} de Vaucouleurs G., 1948, AnAp, 11, 247
\bibitem[\protect\citeauthoryear{}{}]{b1} Dickinson M. et al. 2003, The Mass of Galaxies at Low and High Redshift: Proceedings of the European Southern Observatory, pp. 324
\bibitem[\protect\citeauthoryear{}{}]{b1} Dom\'inguez-Palmero L. \& Balcells M., 2008, A\&A, 489, 1003
\bibitem[\protect\citeauthoryear{}{}]{b1} Eliche-Moral M. C. et al. 2010, A\&A, 519, 55
\bibitem[\protect\citeauthoryear{}{}]{b1} Feldmann R., Carollo C. M., Mayer L. 2011, ApJ, 736, 88
\bibitem[\protect\citeauthoryear{}{}]{b1} Freeman K.C.1970, ApJ, 160, 811
\bibitem[\protect\citeauthoryear{}{}]{b1} Giavalisco M., et al., 2004, ApJ, 600, L93
\bibitem[\protect\citeauthoryear{}{}]{b1} H\"au\ss ler et al. 2007, ApJS, 172, 615
\bibitem[\protect\citeauthoryear{}{}]{b1} Hopkins P. F., Bundy K., Murray N., Quataert E., Lauer T. R., Ma C.-P., 2009, MNRAS, 398, 898
\bibitem[\protect\citeauthoryear{}{}]{b1} Huertas-Company M., Rouan D., Tasca L., Soucail G., Le F\`{e}vre, O., 2008, A\&A,, 478, 971
\bibitem[\protect\citeauthoryear{}{}]{b1} Huertas-Company, M. et al. 2011, A\&A, 525, 157
\bibitem[\protect\citeauthoryear{}{}]{b1} Ilbert O. et al. 2010, ApJ, 709, 644
\bibitem[\protect\citeauthoryear{}{}]{b1} Kajisawa M. et al. 2009, ApJ, 702, 1393
\bibitem[\protect\citeauthoryear{}{}]{b1} Kaviraj, S. 2010, MNRAS, 406, 382
\bibitem[\protect\citeauthoryear{}{}]{b1} Kere\v s D., Katz N., Weinberg D. H., Dav\'{e} R., 2005, MNRAS, 363, 2
\bibitem[\protect\citeauthoryear{}{}]{b1} Khochfar S. \& Silk J. 2006, ApJ, 648, L21
\bibitem[\protect\citeauthoryear{}{}]{b1} Khochfar S. \& Silk J. 2009, MNRAS, 397, 506
\bibitem[\protect\citeauthoryear{}{}]{b1} Law D. R. et al., 2012, ApJ, 745, 85
\bibitem[\protect\citeauthoryear{}{}]{b1} Lintott, C. et al. 2011, MNRAS, 410, 166
\bibitem[\protect\citeauthoryear{}{}]{b1} L\'{o}pez-Sanjuan C. et al. 2010, A\&A, 518, 20
\bibitem[\protect\citeauthoryear{}{}]{b1} L\'{o}pez-Sanjuan C. et al. 2011, A\&A, 530, 20
\bibitem[\protect\citeauthoryear{}{}]{b1} L\'{o}pez-Sanjuan C. et al. 2012, arXiv: 1202.4674
\bibitem[\protect\citeauthoryear{}{}]{b1} Lotz J. M. et al. 2008, ApJ, 672, 177
\bibitem[\protect\citeauthoryear{}{}]{b1} Marchesini D. et al. 2007, ApJ, 656, 42
\bibitem[\protect\citeauthoryear{}{}]{b1} Marchesini D. et al. 2009, ApJ, 701, 1765
\bibitem[\protect\citeauthoryear{}{}]{b1} M\'armol-Queralt\'o, E. et al. 2012, MNRAS, 422, 2187
\bibitem[\protect\citeauthoryear{}{}]{b1} McLure R. J. et al. 2012, MNRAS submitted \& arXiv:1205.4058
\bibitem[\protect\citeauthoryear{}{}]{b1} McGrath, Elizabeth J.; Stockton, Alan; Canalizo, Gabriela; Iye, Masanori; Maihara, Toshinori, 2008, ApJ, 682, 303
\bibitem[\protect\citeauthoryear{}{}]{b1} McIntosh D. et al., 2005, ApJ, 632, 191
\bibitem[\protect\citeauthoryear{}{}]{b1} M\"ollenhoff, C.; Popescu, C. C.; Tuffs, R. J. 2006, A\&A, 456, 941
\bibitem[\protect\citeauthoryear{}{}]{b1} Mortlock A. et al. 2011, MNRAS, 413, 2845
\bibitem[\protect\citeauthoryear{}{}]{b1} Naab T., Johansson P. H., Ostriker J. P., ApJ, 2009, 699, L178
\bibitem[\protect\citeauthoryear{}{}]{b1} Oesch P. et al., 2010, ApJ, 714, L470
\bibitem[\protect\citeauthoryear{}{}]{b1} Oser L. et al. 2010, ApJ, 725, 2312
\bibitem[\protect\citeauthoryear{}{}]{b1} Papovich C. et al. 2006, AJ, 132, 231
\bibitem[\protect\citeauthoryear{}{}]{b1} Peng C. Y. et al. 2010, AJ, 139, 2097
\bibitem[\protect\citeauthoryear{}{}]{b1} P\'{e}rez-Gonz\'{a}lez P. G., et al., 2008a, ApJ, 675, 234
\bibitem[\protect\citeauthoryear{}{}]{b1} P\'{e}rez-Gonz\'{a}lez P. G., Trujillo, I. et al., 2008b, ApJ, 687, 50
\bibitem[\protect\citeauthoryear{}{}]{b1} Prescott M., Baldry, Ivan K., James P. A., 2009, MNRAS, 397, 90
\bibitem[\protect\citeauthoryear{}{}]{b1} Popesso P. et al., 2009, A\&A, 494, 443
\bibitem[\protect\citeauthoryear{}{}]{b1} Quilis V. \& Trujillo I., ApJ, 2012, 752, L19
\bibitem[\protect\citeauthoryear{}{}]{b1} Ravindranath S. et al. 2004, ApJ, 604, L9
\bibitem[\protect\citeauthoryear{}{}]{b1} Ricciardelli E. et al. 2010, MNRAS, 406, 230
\bibitem[\protect\citeauthoryear{}{}]{b1} Rudnick G. et al. 2003, ApJ, 599, 847
\bibitem[\protect\citeauthoryear{}{}]{b1} Ryden B. S., Forbes D. A. \& Terlevich A. I. 2001, MNRAS, 326, 1141
\bibitem[\protect\citeauthoryear{}{}]{b1} Ryden B. S. 2004, ApJ, 601, 214
\bibitem[\protect\citeauthoryear{}{}]{b1} S\'ersic J.-L., 1968, Atlas de Galaxias Australes (C\'ordoba: Observatorio Astron\'omico)
\bibitem[\protect\citeauthoryear{}{}]{b1} Shankar F. et al. 2011, arXiv: 1105.6043
\bibitem[\protect\citeauthoryear{}{}]{b1} Shen S. et al., 2003, MNRAS, 343, 978
\bibitem[\protect\citeauthoryear{}{}]{b1} Toft S. et al. 2007, ApJ, 671, 285
\bibitem[\protect\citeauthoryear{}{}]{b1} Trujillo I., F$\ddot{o}$rster Schreiber N. M., Rudnick G., et al., 2006a, ApJ, 650, 18 
\bibitem[\protect\citeauthoryear{}{}]{b1} Trujillo I., Conselice C. J., Bundy K., Cooper M. C., Eisenhardt P., Ellis R., 2007, MNRAS, 382, 109
\bibitem[\protect\citeauthoryear{}{}]{b1} Trujillo I., Ferreras I. \& De la Rosa I., 2011, MNRAS, 415, 3903
\bibitem[\protect\citeauthoryear{}{}]{b1} Viero M. et al. 2012, MNRAS, 421, 2161
\bibitem[\protect\citeauthoryear{}{}]{b1} van der Wel A. et al. 2011, ApJ, 730, 38
\bibitem[\protect\citeauthoryear{}{}]{b1} van Dokkum P. G. et al. 2010, ApJ, 709, 1018
\bibitem[\protect\citeauthoryear{}{}]{b1} Weinzirl T. et al. 2011, ApJ, 743, 87
\bibitem[\protect\citeauthoryear{}{}]{b1} Wuyts S. et al. 2010, ApJ, 722, 1666
\bibitem[\protect\citeauthoryear{}{}]{b1} Yan H. et al. 2004, ApJ, 616, 63
\end{thebibliography}
\end{document}